%% file: main.tex
\documentclass[sigconf]{acmart}

\usepackage{caption}
\usepackage{subcaption}     
\usepackage{xcolor}
\usepackage{multirow}
\usepackage{threeparttable}
\usepackage{enumitem}
\usepackage{siunitx}
\usepackage{pifont}
\usepackage{algorithm}
\usepackage{algpseudocode}

\AtBeginDocument{%
  }

\copyrightyear{2025}
\acmYear{2025}
\setcopyright{acmlicensed}\acmConference[MM '25]{Proceedings of the 33rd ACM International Conference on Multimedia}{October 27--31, 2025}{Dublin, Ireland}
\acmBooktitle{Proceedings of the 33rd ACM International Conference on Multimedia (MM '25), October 27--31, 2025, Dublin, Ireland}
\acmDOI{10.1145/3746027.3755629}
\acmISBN{979-8-4007-2035-2/2025/10}

\acmSubmissionID{4903}

\begin{document}


\title{Enkidu: Universal Frequential Perturbation for Real-Time Audio Privacy Protection against Voice Deepfakes}


\author{Zhou Feng}
\affiliation{
    \institution{College of Computer Science and Technology, Zhejiang University}
    \city{Hangzhou}
    \country{China}
}
\email{zhou.feng@zju.edu.cn}

\author{Jiahao Chen}
\affiliation{
    \institution{College of Computer Science and Technology, Zhejiang University}
    \city{Hangzhou}
    \country{China}
}
\email{xaddwell@zju.edu.cn}

\author{Chunyi Zhou}
\authornote{Corresponding author.}
\affiliation{
    \institution{College of Computer Science and Technology, Zhejiang University}
    \city{Hangzhou}
    \country{China}
}
\email{zhouchunyi@zju.edu.cn}

\author{Yuwen Pu}
\affiliation{
    \institution{School of Big Data \& Software Engineering, Chongqing University}
    \city{Chongqing}
    \country{China}
}
\email{yw.pu@cqu.edu.cn}

\author{Qingming Li}
\affiliation{
    \institution{College of Computer Science and Technology, Zhejiang University}
    \city{Hangzhou}
    \country{China}
}
\email{liqm@zju.edu.cn}

\author{Tianyu Du}
\affiliation{
    \institution{College of Computer Science and Technology, Zhejiang University}
    \city{Hangzhou}
    \country{China}
}
\email{zjradty@zju.edu.cn}

\author{Shouling Ji}
\affiliation{
    \institution{College of Computer Science and Technology, Zhejiang University}
    \city{Hangzhou}
    \country{China}
}
\email{sji@zju.edu.cn}

\renewcommand{\shortauthors}{F. Zhou, J. Chen, C. Zhou et al.}

\input{sections/abstract}



\received{11 April 2025}
\received[revised]{15 September 2025}
\received[accepted]{6 July 2025}

\maketitle

\input{sections/introduction}

\input{sections/background}

\input{sections/threat_model}

\input{sections/method}

\input{sections/evaluation}

\input{sections/conclusion}

\begin{acks}
This work was partly supported by the National Key Research and Development Program of China under No. 2022YFB3102100, NSFC under No. U244120033, U24A20336, 62172243, 62402425 and 62402418, the China Postdoctoral Science Foundation under No. 2024M762829, the Zhejiang Provincial Natural Science Foundation under No. LD24F020002, the "Pioneer and Leading Goose" R\&D Program of Zhejiang under 2025C01082, 2025C02033 and 2025C02263, and the Zhejiang Provincial Priority-Funded Postdoctoral Research Project under No. ZJ2024001.
\end{acks}

\input{main.bbl}


\input{sections/appendix}

\end{document}

%% file: sections/abstract.tex
\begin{abstract}
The rise of advanced voice deepfake technologies has raised serious concerns over user audio privacy, as malicious actors increasingly exploit publicly available voice data to generate convincing fake audio for malicious purposes such as identity theft, financial fraud and misinformation campaigns. 
While existing defense methods offer partial protection, they suffer from critical limitations, including \textbf{weak adaptability} to unseen user data, \textbf{poor scalability} to long audio, \textbf{regid reliance on white-box knowledge} and \textbf{high computational and temporal costs} to encryption process. 
Therefore, to defend against personalized voice deepfake threats, we propose \textit{Enkidu}, a novel user-oriented privacy-preserving framework that leverages universal frequential perturbations generated through black-box knowledge and few-shot training on a small amount of user samples.
These high-malleablity frequency-domain noise patches enable real-time, lightweight protection with strong generalization across variable-length audio and robust resistance against voice deepfake attacks—all while preserving high perceptual and intelligible audio quality.
Notably, \textit{Enkidu} achieves over \textbf{50–200$\times$ processing memory efficiency} (requiring only \textbf{0.004 GB}) and over \textbf{3–7000$\times$ runtime efficiency} (real-time coefficient as low as \textbf{0.004}) compared to six SOTA countermeasures.
Extensive experiments across six mainstream Text-to-Speech (TTS) models and five cutting-edge Automated Speaker Verification (ASV) models demonstrate the effectiveness, transferability, and practicality of \textit{Enkidu} in defending against voice deepfakes and adaptive attacks. Our code is currently available\footnote{\url{https://github.com/NoobCodeNameless/Enkidu}}.
\end{abstract}

%% file: sections/introduction.tex


\section{Introduction}
Speech synthesis technologies, driven predominantly by deep learning breakthroughs, have witnessed remarkable advancements in recent years~\cite{Tan21Survey, Zhang23AudioDiffusionSurvey}. Voice deepfake can now generate synthetic speech indistinguishable from real human voices via mighty Text-to-Speech (TTS) systems, significantly enhancing applications ranging from personalized virtual assistants to automated narration and entertainment industries~\cite{Kepuska18VPAs, Yadav23VoiceVPA, MurfAIVoiceMedia2024}. Powered by sophisticated neural architectures and massive datasets, these technologies produce remarkably realistic audio outputs, which have rapidly proliferated and become widely accessible to the general public~\cite{Amezaga22VoiceDeepfake}.

\begin{figure}[]
  \centering
  \includegraphics[width=0.92\linewidth]{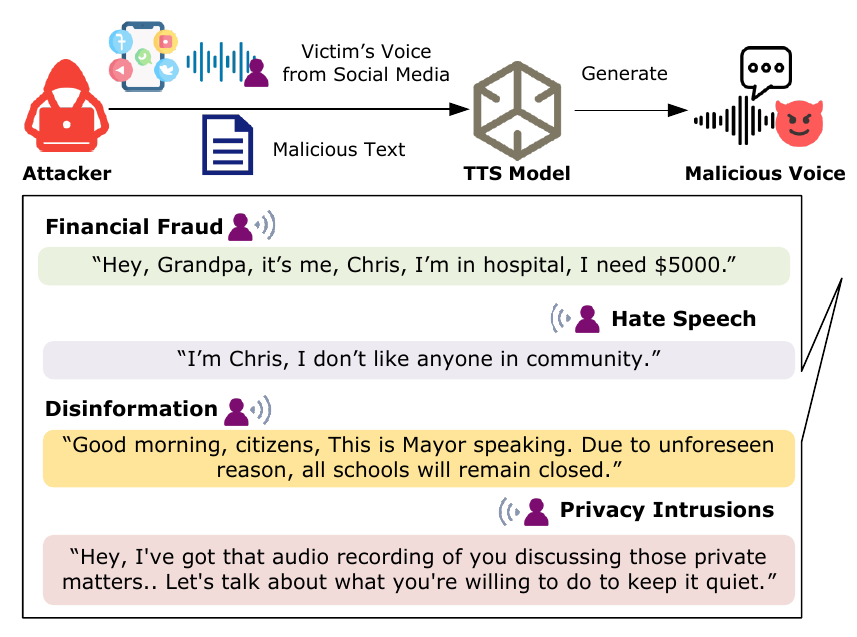}
  \caption{The real-life threat of voice deepfake misusage.}
  \label{fig:scenario}
\end{figure}

\input{table/comparision}
Despite these notable advancements, the proliferation of sophisticated speech synthesis systems has simultaneously given rise to severe privacy concerns, particularly for common users~\cite{NDTV2024AIScamsIndia, Alspach2024AudioDeepfake}. Specifically, individuals frequently share their audio samples publicly on social media and other online platforms (e.g., Spotify, SoundCloud and YouTube), unknowingly exposing their biometric voice characteristics to significant misuse. As illustrated in Figure~\ref{fig:scenario}, such openly accessible voice data may be exploited by malicious entities to craft convincing deepfake audio, posing real-world threats such as financial fraud~\cite{Vakulov2025DeepfakeScam}, hate speech~\cite{Finley2024DeepfakePrincipal}, disinformation~\cite{Cohen2024DisinfoDeepfake} and privacy intrusions~\cite{MizterAB2023DeepfakePrivacy}. Common users exhibit significant vulnerability due to limited access to effective privacy-preserving mechanisms and insufficient risk awareness.

Several methodologies have emerged to counteract these privacy threats. Detection technologies~\cite{Shiota15Liveness, Shang18VoiceLiveness, Zhang17VoiceGesture, Ahmed20Void, Blue22DeepfakeDetect} were initially developed to identify suspicious deepfake audio, but researchers increasingly emphasize proactive defense methods to prevent audio deepfake at their source. 
AntiFake~\cite{Yu23AntiFake} adversarially optimizes audio samples to mislead speech synthesis models into generating incorrect speaker identities, while POP~\cite{Zhang24LAMPS} embeds imperceptible perturbations optimized for reconstruction loss, rendering the protected samples unlearnable by TTS models. Additionally, V-CLOAK~\cite{Deng23VCloak} achieves speaker anonymization via a one-shot adversarial generative approach that preserves intelligibility and timbre.

Nevertheless, current proactive defense techniques~\cite{Yu23AntiFake, Zhang24LAMPS, Deng23VCloak} universally face critical limitations, including \textbf{limited adaptability} to unseen user data, \textbf{inability to handle long-duration audio} effectively, \textbf{unsustainable reliance} on white-box knowledge (accessibility to voice deepfake models) and \textbf{prohibitive temporal and computational costs}. Thus, it remains essential to develop a privacy-preserving approach that is simultaneously effective, scalable, practical and efficient while preserving acoustic fidelity.

To address these pressing issues, we propose \textit{Enkidu}, a novel user-oriented audio privacy-preserving framework utilizing Universal Frequential Perturbations (UFP) against voice deepfake threats targeting specific users. To succinctly highlight the capabilities and practical strengths of our proposed approach, consider the overview presented in Table~\ref{tab:enkidu_comparison} along with the following discussions:





\noindent\textbf{Q1:} \textit{Is there a method that ensures audio real-time privacy protection?}

\noindent\textbf{A1:} \textbf{Yes}. By generating user-specific UFP patches in advance through few-shot training, our method enables real-time attachment to any user audio samples.

\noindent\textbf{Q2:} \textit{Can voice privacy be effectively protected on resource-constrained devices with low computational overhead?}

\noindent\textbf{A2:}\textbf{Indeed}. By significantly reducing GPU consumption during the noise attachment, our UFP-based method is highly suitable for deployment even on edge devices with limited computing resources.

\noindent\textbf{Q3:} \textit{Can audio of arbitrary length be protected consistently without compromising holistic quality or performance?}  

\noindent\textbf{A3:} \textbf{Absolutely}. Our UFP seamlessly accommodate audio samples of any duration, ensuring robust and consistent privacy protection. 

\noindent\textbf{Q4:} \textit{Given these capabilities, can such a solution preserve excellent acoustic clarity for human listeners and intelligibility for automatic speech recognition (ASR) systems?}  

\noindent\textbf{A4:} \textbf{Precisely}. Leveraging psychoacoustic principles, our method ensures high audio quality, making noise imperceptible to humans while preserving intelligibility for ASRs.

Our contributions of this paper can be summarized as follows:
\begin{itemize}
    \item We propose a novel user-oriented framework for proactive audio privacy preservation, enabling effective, scalable, practical and efficient protection against voice deepfake attacks with low perceptual distortion even under black-box settings.
    \item We introduce UFP optimized through few-shot training approach, ensuring real-time deployment, low computational overhead, and scalability across audio of varying lengths.
    \item  We conduct extensive evaluations across multiple TTS and ASV models, demonstrating strong privacy-preserving performance, robustness under adaptive attacks, and efficiency under real-time and resource-constrained settings. Ablation studies further validate the effectiveness and generalizability of our design choices.
\end{itemize}

%% file: table/comparision.tex
\begin{table*}[t]
\tabcolsep=0.24cm
\centering
\begin{threeparttable}
\caption{Comparison with existing audio privacy-preserving methods. \ding{52}/\ding{56} indicates whether the method satisfies the corresponding property.}
\label{tab:enkidu_comparison}
\begin{tabular}{ccccccccc}
\toprule
\multirow{2}{*}{\textbf{Methods}} & \multirow{2}{*}{\textbf{Type}} & \multirow{2}{*}{\textbf{Knowledge}} & \multicolumn{2}{c}{\textbf{Effectiveness}} & \multicolumn{2}{c}{\textbf{Universality}} & \multicolumn{2}{c}{\textbf{Efficiency}} \\
\cline{4-9}
& & & Robust. & Qual. & Trans. & ALA & Mem. (GB) $\downarrow$ & RTC $\downarrow$ \\
\midrule
AntiFake~\cite{Yu23AntiFake} & VP & White-box & \ding{52} & \ding{52} & \ding{56} & \ding{56} & $\approx 6-8$ & 28.2 \\
VoiceGuard~\cite{Li23VoiceGuard} & VP & White-box & \ding{52} & \ding{52} & \ding{56} & \ding{56} & $\approx 3.5-5.5$ & 0.45 \\
VSMask~\cite{Wang23VSMask} & VP & White-box & \ding{52} & \ding{52} & \ding{56} & \ding{56} & $\approx 0.2-0.3$ & 0.213 \\
POP~\cite{Zhang24LAMPS} & VP & White-box/Black-box & \ding{52} & \ding{52} & \ding{52} & \ding{56} & $\approx 3-4$ & 1.855 \\
SVTaM~\cite{Yao24SVT} & SA & White-box & \ding{56} & \ding{52} & \ding{56} & \ding{56} &$\approx 5-7.5$ & 0.2 \\
V-CLOAK~\cite{Deng23VCloak} & SA & White-box & \ding{56} & \ding{52} & \ding{56} & \ding{52} & $\approx 2.5-3.5$ & 0.011 \\
\textbf{Ours (Enkidu)} & \textbf{Universal VP} & \textbf{Black-box} & \ding{52} & \ding{52} & \ding{52} & \ding{52} & \textbf{0.004} & \textbf{0.004} \\
\bottomrule
\end{tabular}
\begin{tablenotes}
\small
\item \textbf{VP}: Voice-based Perturbation; \textbf{SA}: Speaker Anonymization;  \textbf{Robust.}: Voice Deepfake Robustness; \textbf{Qual.}: Perceptual Quality; \textbf{Trans.}: Transferability, the ability of the perturbation to generalize across unseen audio samples; \textbf{ALA}: Audio-Length Agnostic; \textbf{RTC}: Real-Time Coefficient, the ratio between the processing time and the audio length (lower is better).
\end{tablenotes}
\end{threeparttable}
\end{table*}

%% file: sections/background.tex

\section{Preliminaries}

\subsection{Speaker \& Speech Recognition Systems}
Automatic Speaker Verification (ASV) systems are designed to authenticate or verify a speaker's identity based solely on voice data. Modern ASV frameworks typically begin by converting raw audio waveforms into standardized acoustic representations. These acoustic features are then fed into neural network-based embedding extractors such as X-Vector networks~\cite{Snyder18XVector}, ECAPA-TDNN~\cite{Desplanques20ECAPA}, or ResNet-based architectures~\cite{He16ResNet, Chen23EnhancedRes2Net} to produce fixed-dimensional embeddings, often referred to as speaker embeddings or voiceprints. These embeddings ideally encapsulate distinctive and stable speaker-specific characteristics, minimizing variability due to environmental noise, recording conditions, or linguistic content. During verification, embeddings from test samples are compared against reference embeddings using similarity measures like cosine similarity. A pre-defined threshold, is then used to decide if two embeddings represent the same speaker.

Automatic Speech Recognition (ASR), in contrast, aims to transcribe spoken language into textual representations accurately. Recent advancements in deep learning have significantly enhanced ASR performance. Modern ASR systems frequently adopt sophisticated neural architectures such as Recurrent Neural Networks (RNNs) with Long Short-Term Memory (LSTM) units, Convolutional Neural Networks (CNNs), and, more prominently, Transformer-based models like Whisper~\cite{Radford23Whisper} and wav2vec2~\cite{Baevski20Wav2vec2}. 

\subsection{Voice Deepfake System}
Voice deepfake technologies have significantly advanced alongside rapid developments in deep learning, with Text-to-Speech (TTS) systems serving as their primary enabling tool. Early voice synthesis methods, such as concatenative synthesis~\cite{Olive77DyadicUnits, Moulines90PSOLA, Sagisaka92ATR, Hunt96UnitSelection}, relied heavily on stitching together pre-recorded speech segments. Later, statistical parametric synthesis emerged, leveraging machine learning frameworks like Hidden Markov Models (HMMs)~\cite{Starner95ASL} to model acoustic parameters explicitly. Modern voice deepfake systems are predominantly powered by neural-based TTS methods, exemplified by WaveNet~\cite{Oord16WaveNet}, Tacotron~\cite{Wang17Tacotron}, FastSpeech~\cite{Ren19FastSpeech}, and their derivatives~\cite{Ren21FastSpeech2, Shen18TacoTron2}, capable of producing speech that is increasingly natural and expressive.
Though sharing similarities, subtle conceptual differences exist between TTS and voice conversion (VC) systems. VC typically involves modifying existing speech to change speaker-specific attributes while maintaining linguistic content~\cite{Sisman2020VoiceConversion}. Conversely, TTS system synthesize speech directly from textual input, simultaneously generating both linguistic and acoustic information. Nevertheless, advancements in neural network methodologies have increasingly integrated aspects of these two approaches, resulting in less distinct boundaries.

Contemporary SOTA TTS systems primarily employ end-to-end neural architectures such as Tacotron2~\cite{Shen18TacoTron2}, FastPitch~\cite{Lancucki21FastPitch}, and YourTTS~\cite{Casanova22YourTTS}, offering significant improvements in naturalness, intelligibility, and expressivity. These models typically utilize en-coder-decoder architectures, Transformer-based~\cite{Li19TransformerTTS} self-attention mechanisms, and vocoders that convert predicted spectrograms into waveforms, significantly narrowing the gap between synthesized and natural human speech, which demonstrate exceptional effectiveness even under few-shot or zero-shot learning scenarios. These systems can synthesize realistic and personalized speech using a relatively small number of voice samples even in one, greatly enhancing TTS flexibility and applicability.


\subsection{Anti-Voice Deepfake Defenses}

With the rapid development of voice deepfake technologies, the risk of malicious misuse has increased significantly, prompting extensive research into countermeasures. Existing anti-voice-deepfake strategies can be broadly categorized into two paradigms: synthesized audio detection and proactive defenses. 

Synthesized audio detection primarily targets two key aspects: (1) \textit{liveness detection}, which leverages physical properties of real-world recording conditions~\cite{Shiota15Liveness, Shang18VoiceLiveness, Zhang17VoiceGesture}, and (2) \textit{signal artifact analysis}, which detects subtle artifacts introduced by synthesis pipelines~\cite{Ahmed20Void, Blue22DeepfakeDetect}. Although these approaches initially achieved promising results, modern TTS systems have advanced to the point of accurately simulating emotional expression and environmental noise~\cite{Chen24VALLE2}, significantly narrowing the perceptual gap and challenging the robustness of detection-based methods.

However, even when detection methods succeed in identifying synthetic speech, they do so reactively—after the user’s voice features may have already been exploited. In contrast, proactive defenses take a preventative stance, addressing potential threats at their origin. These approaches fall broadly into two categories: speaker anonymization and voice-based perturbation.

Speaker anonymization attempts to obfuscate or replace speaker identity within the audio, typically through adversarial transformation or VC techniques. For instance, Fang et al.~\cite{Fang19SSW} propose a method that manipulates x-vector representations to derive anonym-ized pseudo speaker identities via multiple combinations. To address the inconsistency and instability of x-vector transformations~\cite{Panariello23VocoderDrift, Panariello23VocoderDriftIS}, Panariello et al.~\cite{Panariello24ICASSP} introduce a neural codec-based anonymization technique that generates high-quality anonymous speech, while Yao et al.~\cite{Yao24SVT} propose the SVTaM framework, which avoids the limitations of traditional x-vector averaging and external speaker pools. Despite their effectiveness, these methods often struggle with real-time processing and variable-length inputs. To address these limitations, V-CLOAK~\cite{Deng23VCloak} proposes a one-shot anonymization model based on Wave-U-Net~\cite{Stoller18WaveUNet}, supporting real-time, arbitrary-length audio anonymization. While speaker anonymization provides a strong layer of privacy protection, its robustness against voice cloning models and the transferability of its transformations remain underexplored.


Voice-based perturbation methods instead aim to subtly modify the original audio signal to degrade the performance of voice cloning models. For example, AntiFake~\cite{Yu23AntiFake} generates adversarial perturbations tailored to individual utterances to mislead TTS models. VoiceGuard~\cite{Li23VoiceGuard} improves stealthiness by applying perturbations directly in the time domain, using a psychoacoustic masking model to conceal them. VSMask~\cite{Wang23VSMask} extends this idea into real-time settings by injecting perturbations into live speech streams. While these methods offer strong trade-offs between robustness and perceptual quality, their perturbations are generally sample-specific and lack transferability, often relying on white-box access or assumptions that are sort of impractical in real-world deployments. To overcome this, POP~\cite{Zhang24LAMPS} proposes a universal, imperceptible perturbation patch optimized with reconstruction loss, which renders protected voice samples unlearnable by TTS models and exhibits partial transferability across samples. However, its effectiveness remains limited when transitioning from white-box to black-box settings. Nonetheless, achieving holistic transferability, black-box compatibility, and robustness to arbitrary-length audio remains an open challenge in perturbation-based defenses.

%% file: sections/threat_model.tex
\begin{figure}[]
  \centering
  \includegraphics[width=0.45\textwidth]{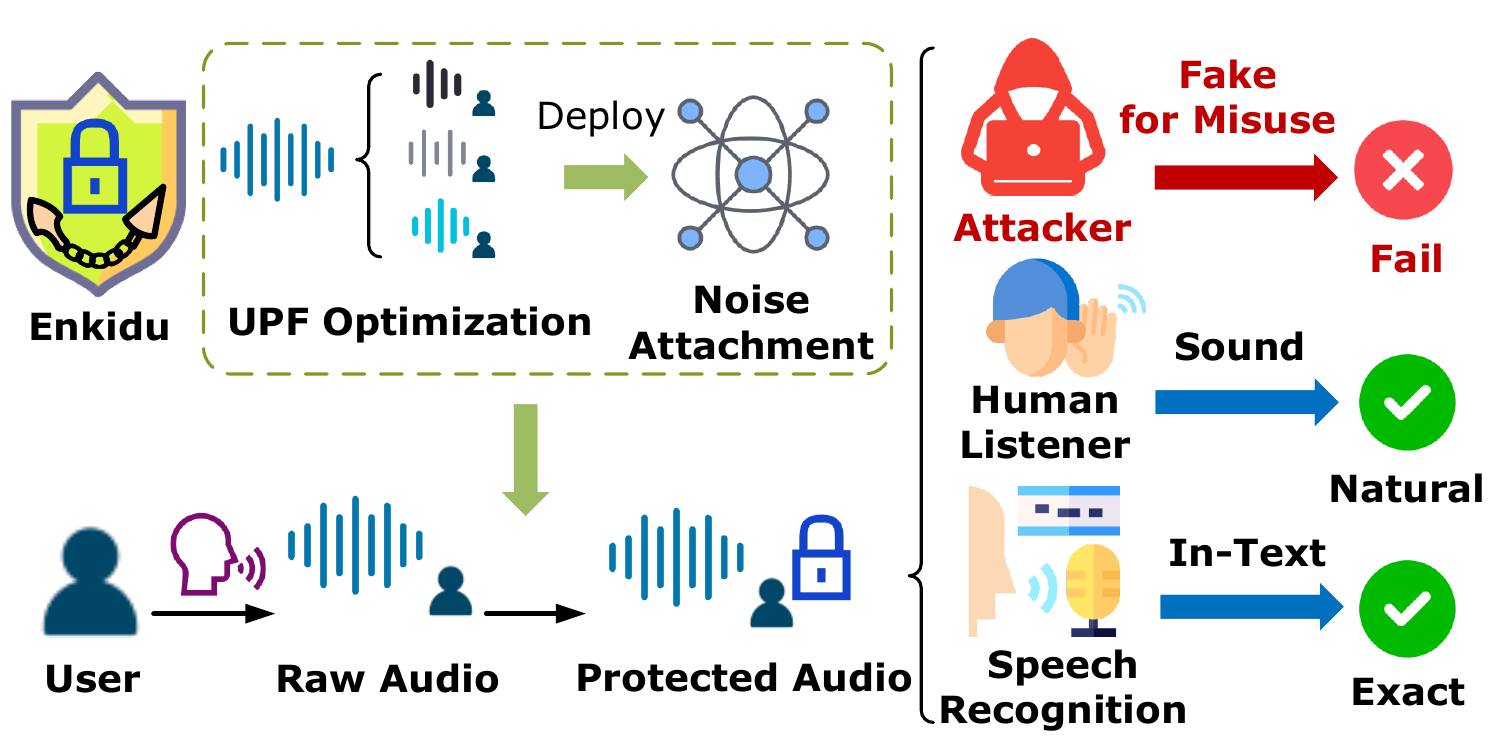}
  \caption{Threat model. The \textit{Enkidu} generates optimized UFP and attaches it to user audio in real time. The protected audio maintains naturalness for human listeners and transcription accuracy, while degrading performance of malicious TTS-based voice mimicry, thus preventing misuse.}
  \Description{The threat model of the paper.}
  \label{fig:threat_model}
\end{figure}

\section{Threat Model}
In this section, we outline the threat model by describing the motivations behind potential attacks and specifying the assumptions about the attacker’s capabilities and objectives.
\subsection{Attacker Motivation}
Attackers leveraging voice deepfake technologies are primarily motivated by the possibility of maliciously generating deceptive audio content that convincingly mimics a victim’s voice. Such attacks may aim at perpetrating fraud, misinformation, or social engineering schemes by exploiting trust placed in familiar voices. 


    
    


\subsection{Attacker Assumption}
\label{subsec:attacker}

\textbf{Attacker’s Goal.}
The attacker’s accessible data is limited to voice samples published by the victim online. Since the attacker has no access to the victim’s real-time or high-fidelity voice characteristics, their initial goal is to infer a generalizable voice representation from the limited public recordings. The ultimate objective is to synthesize audio samples that closely approximate the victim’s authentic voice, not just in acoustic similarity, but also in terms of identity verification metrics. These synthetic voices must be realistic enough to deceive both ASV systems and human listeners, enabling successful impersonation, manipulation, or identity fraud in downstream applications.

\noindent\textbf{Attacker’s Capability.}
We assume the attacker has the capability to collect voice samples of the victim from publicly available sources (e.g., interviews, social media posts, or podcasts). After obtaining the voice data, the attacker may either: (1) directly input the raw audio and malicious text into a TTS system, or (2) preprocess the audio data (e.g., resampling, normalizing, or denoising) before feeding it into the synthesis model. Leveraging modern TTS systems, the attacker can synthesize arbitrary speech that imitates the victim’s voice, as illustrated in Figure~\ref{fig:threat_model}. These generated audios are crafted to be perceptually indistinguishable from the victim's authentic voice for both human listeners and ASV systems.

\subsection{Defender Assumption}
\label{subsec:defense}

\textbf{Defender’s Goal.}
The defender’s primary objective is to publish the user's audio samples in a manner that preserves usability and naturalness while preventing malicious voice antifakes attacks. To achieve this, the defender introduces well-optimized universal perturbations to the audio before it is made public, aiming to disrupt potential misuse by TTS or other voice deepfake systems. 
Therefore, the defender targets two levels of protection. 

\textit{Shallow Protection.} The perturbed audio, when collected and used by the attacker, exhibits unlearnable or misleading voice features for voice deepfake systems. This causes a misalignment between the synthesized audio and the original speaker identity, leading to low consistency across features extracted from the perturbed and synthesized audios. 

\textit{Deep Protection.} The perturbations not only confuse feature learning but also fundamentally mislead the synthesis process. As a result, the generated audio shows a strong identity mismatch compared to the user’s original unperturbed voice, offering deeper and more robust protection against impersonation attempts.




\noindent\textbf{Defender’s Capability.}
We assume that the defender has the ability to control the user's audio content before it is published on public platforms, such as social media. Additionally, the defender operates under black-box setting and must efficiently generate a user-specific, universal perturbation that can be applied across different audio samples. This perturbation should be lightweight enough for real-time application and robust enough to protect the user's identity in diverse audio contexts.

%% file: sections/method.tex

\section{Methodology}

\subsection{Problem Formulation}
\label{subsec:problem}
Given the voice sample $x$ to be published by user $u$ (victim), and the corresponding set $\mathcal{D}_u = \{x_1, \ldots, x_N\}$ with $N$ samples. As mentioned above, the defenders attempt to apply the defensive perturbation $\delta$ to the input sample $x$ with function $\tilde{x} = \mathcal{F}(x, \delta)$ to obtain the protected voice sample $\tilde{x}$. The potential attackers that collect the victim's voice sample, use TTS models to synthesize a speech sample $\mathcal{G}(\tilde{x}, t)$ conditioned on input speech $\tilde{x}$ and target text $t$. Therefore, the optimization goal is to find a $\delta$ such that satisfies:

\begin{equation}
    \min_{\delta} \ \mathbb{E}_{x_i \sim \mathcal{D}_u} \left[\mathcal{H}(\tilde{x}_i) -\mathrm{SV}(\tilde{x}_i, \mathcal{G}(\tilde{x}_i, t); \tau)\right],
\label{eq:opt_goal}
\end{equation}
where $\mathcal{H}(\cdot)$ stands for the perceptual metric function that evaluates the naturalness and intelligibility (the higher, the better) of a speech sample. Additionally, $\mathrm{SV}(x_1, x_2; \tau)$ denotes the ASV decision function that determines whether the given two samples are from speakers of the same identity, based on a threshold $\tau$:
    \begin{equation}
        \mathrm{SV}(x_1, x_2; \tau) = \mathbb{I} \left( \mathrm{Dist}(\mathcal{E}(x_1), \mathcal{E}(x_2)) \ge \tau \right),
        \label{eq:sv}
    \end{equation}
where $\mathbb{I}(\cdot)$ is the indicator function and $\mathcal{E}(x)$ is a voice feature encoder that maps a speech sample $x$ to its embedding that represents one's voice identity feature. Next, the similarity of the voice features is measured by $\mathrm{Dist}(\cdot, \cdot)$.

The two terms in Equation~\ref{eq:opt_goal} aim to reduce the success rate of voice cloning, respectively, by ensuring that even cloned voices generated from perturbed inputs fail to reach a generalization among perturbed samples and reach the voice feature of the real identity. The last term ensures that the perturbed audio remains perceptually natural and intelligible. However, we shall also emphasize that Equation~\ref{eq:opt_goal} relies on the accessibility to the TTS model of the attackers, which is not practical. In the following content, we relax this assumption and propose a more practical black-box optimization strategy that does not require access to the TTS model, leveraging surrogate models and transferable representations to maintain the effectiveness of UFP.


\subsection{Overview of Enkidu}

To defend against TTS-based voice deepfake attacks, we propose \textit{Enkidu}, a novel, user-oriented audio privacy-preserving framework by introducing an optimized UFP into user audio. The design of \textit{Enkidu} is guided by two primary objectives:

\begin{itemize}
    \item \textbf{Ensuring the identity protection of user speech}: The UFP should effectively degrade speaker-specific representations extracted by TTS systems, while preserving the perceptual quality of the audio to remain natural and intelligible to human listeners.

    \item \textbf{Supporting Real-time, low-cost encryption across diverse scenarios}: The system should flexibly process variable-length audio and run efficiently on resource-limited devices for low-latency, scalable deployment.

\end{itemize}


\textit{Enkidu} adopts a frequency-domain perturbation strategy based on frame-wise tiling: a compact UFP is adaptively aligned with the spectrogram of user audio in either piled or cropped patches, enabling efficient and length-agnostic application. Accordingly, \textit{Enkidu} follows a two-stage workflow: first, it optimizes a user-specific UFP using a small set of clean utterances; then, it applies the learned perturbation to arbitrary-length audio via a lightweight alignment module.

\subsection{Method Design}
\label{subsec:method}

\subsubsection{Stage I: UFP Optimization}
\label{subsubsec:optimization}

Since TTS systems primarily rely on specific frequency bands to extract speaker characteristics~\cite{Kim20GlowTTS,Casanova22YourTTS}, we aim to learn a UFP malleable to the spectrogram of user audio, represented by a complex-valued matrix $\delta = \delta_r + j \cdot \delta_i \in \mathbb{C}^{1 \times B \times L_u}$, where $B$ is the number of frequency bins and $L_u$ is the frame length.

\input{table/main_experiment}

The goal is to disrupt speaker embeddings extracted from perturbed speech while maintaining perceptual audio quality. To achieve this, we define two losses. \\
\textbf{Feature Disruption Loss} $\mathcal{L}_{\text{fea}}$: Note that unlike previous works~\cite{Yu23AntiFake}, we do not assume that the defender has access to the details of TTS (e.g., the feature extractor). Instead, we measure the distance between the features of original and perturbed audio using a pre-trained speaker verification encoder $\mathcal{E}(\cdot)$. This is motivated by the observation that speaker verification encoders are often trained to capture speaker identity information, which aligns closely with the feature representations used by TTS models to maintain speaker consistency during synthesis. \\
\textbf{Perception Loss} $\mathcal{L}_{\text{per}}$: Encourages the protected voice sample $\tilde{x}_i = \mathcal{F}(x_i, \delta)$ to remain close to the original in the perceptual domain. Additionally, incorporating frequential perturbation helps preserve the overall spectral structure of speech, allowing the perturbation to remain less perceptible to human listeners while still being effective in disrupting the TTS's internal reconstruction.

The overall training objective is defined as:
\begin{equation}
    \min_{\delta} \; \mathbb{E}_{x_i \in \mathcal{D}_u} \Big[ \mathcal{L}_{\text{fea}}\left(\mathcal{E}(x_i), \mathcal{E}(\tilde{x}_i)\right) + \lambda\cdot \mathcal{L}_{\text{per}}\left(x_i, \tilde{x}_i \right) \Big],
    \label{eq:opt}
\end{equation}
where $\lambda$ is a hyperparameter that balances privacy disruption and perceptual quality. Specifically, we use $\ell_2$ distance for the measurement of both identity feature and perception quality in Equation~\ref{eq:opt}. 

To enhance robustness and imperceptibility, we introduce a frame-wise frequential augmentation strategy tailored to the UFP’s tiling structure, including random temporal shifts and binary masking applied to spectrogram segments during optimization, as described in Section~\ref{subsubsec:tiler}. These augmentations are omitted during deployment, where full-frame tiling is applied without masking. Additionally, time-domain augmentations such as additive noise and temporal jitter are used during training to improve generalization. Implementation details are provided in Algorithms~\ref{alg:ufp-optimization} and Appendix E.


\begin{algorithm}[t]
\caption{User-Oriented UFP Optimization}
\label{alg:ufp-optimization}
\begin{algorithmic}
\Require User Offered Data $\mathcal{D}_t$; Iterations $K$; Frame Length $L$; Noise Level $\eta$;  
\Ensure UFP Perturbation $(\delta_r, \delta_i)$

\Function{UFP}{$\mathcal{D}_t, f$}
\State $\delta_r, \delta_i \sim \mathcal{N}(0, 1)^{1 \times B \times L_u}$ \Comment{$B$: Freq Bins}
\For{$t = 1$ to $K$}
    \For{$x_i \in \mathcal{D}_t$}
        \State $z_i \gets \mathcal{E}(x_i)$ \Comment{$z$: Extract Embedding}
        \State $\tilde{x}_i \gets \textsc{Tiler}(x_i, \delta_r, \delta_i, \eta, L_u, a=1)$
        \State $\tilde{x}_i \gets \text{TemporalAugmentation}(\tilde{x}_i)$
        \State $\tilde{z}_i \gets \mathcal{E}(\tilde{x}_i)$
        \State $\mathcal{L} \gets \mathcal{L}_{\text{fea}}(z_i,\tilde{z}_i) + \lambda \cdot \mathcal{L}_{\text{per}}(x_i) $
        \State $\nabla \gets \nabla + \nabla_{\delta_r, \delta_i}\mathcal{L}$
    \EndFor
    \State Update $\delta_r, \delta_i$ with $\nabla$ using Adam
\EndFor
\State \Return $(\delta_r, \delta_i)$
\EndFunction
\end{algorithmic}
\end{algorithm}

\subsubsection{Stage II: Real-time Encryption via Tiler}
\label{subsubsec:tiler}
The \textit{Tiler} module serves as a dual-purpose encryption component, responsible for both UFP optimization and rapid deployment. It implements the transformation $\tilde{x} = \mathcal{F}(x, \delta)$ by applying a learning or learned UFP to a given audio waveform. The encryption procedure consists of the following steps:

\begin{enumerate}
    \item Convert the input waveform $x$ into its complex-valued spectrogram $S = \text{STFT}(x) \in \mathbb{C}^{1 \times B \times L}$ using Short-Time Fourier Transform (STFT).
    
    \item Smooth the real and imaginary parts of the UFP, $\delta_r$ and $\delta_i$, to suppress abrupt spectral changes. Then compose the complex perturbation $\delta = \delta_r + j \cdot \delta_i \in \mathbb{C}^{1 \times B \times L_u}$.

    \item In both optimization and deployment, the spectrogram $S$ is divided into $\lfloor |S| / L_u \rfloor$ non-overlapping segments aligned with the UFP frame length $L_u$, and the perturbation $\delta$ is tiled across these segments via piling or cropping. In the optimization phase, a temporal shift $\epsilon \in [0, L_u]$ is first applied to $S$, and a binary mask $m$ is initialized over the segments, where each frame is independently selected with probability $(1 - r)$. Perturbation is then applied only to the selected frames, introducing structured sparsity that improves robustness and imperceptibility. In deployment, no shift or masking is applied—the UFP is directly tiled over the spectrogram for full-frame perturbation.

    \item Convert the perturbed spectrogram $\tilde{S}$ back to the time-domain waveform $\tilde{x} = \text{iSTFT}(\tilde{S})$ using the inverse STFT.
\end{enumerate}

Once trained, UFP can be directly applied to any unseen audio via the \textit{Tiler} module. This tiled, frequency-domain perturbation strategy enables low-latency encryption suitable for real-time streaming and deployment on edge devices. Its universal and reusable nature eliminates the need for sample-wise optimization. Further implementation details are provided in Appendix E.

%% file: table/main_experiment.tex
\begin{table*}[t]
  \caption{
    Protection effectiveness of \textit{Enkidu} against various TTS models, evaluated across five ASV backbones. Higher SPR / DPR indicate stronger defense performance. Enkidu maintains high audio quality with a MOS of 3.01±0.07, a STOI of 0.71±0.01 and perfect intelligibility with both CER and WER at 0.00\%±0.00\%.
    }

  \label{tab:main_experiment}
  \centering
  \resizebox{0.90\linewidth}{!}{
  \begin{tabular}{cccccccccccccc}
    \toprule
    \multirow{2}{*}{\textbf{Deepfake Model}} & \multicolumn{2}{c}{\textbf{ECAPA-TDNN}} & \multicolumn{2}{c}{\textbf{X-Vector}} & \multicolumn{2}{c}{\textbf{ResNet}} & \multicolumn{2}{c}{\textbf{ERes2Net}} & \multicolumn{2}{c}{\textbf{Cam++}} & \multicolumn{2}{c}{\textit{Average}} \\
    \cline{2-13}
    & \textbf{SPR} & \textbf{DPR} & \textbf{SPR} & \textbf{DPR} & \textbf{SPR} & \textbf{DPR} & \textbf{SPR} & \textbf{DPR} & \textbf{SPR} & \textbf{DPR} & \textbf{SPR} & \textbf{DPR} \\
    \midrule
    Speedy-Speech~\cite{Vainer20SpeedySpeech}  & 96.55\% & 72.41\% & 75.86\% & 24.14\% & 86.21\% & 96.55\% & 89.66\% & 100.00\% & 89.66\% & 86.21\% & 87.59\% & 75.86\% \\
    FastPitch~\cite{Lancucki21FastPitch}     & 100.00\% & 68.97\% & 72.41\% & 13.79\% & 68.97\% & 79.31\% & 89.66\% & 100.00\% & 72.41\% & 89.66\% & 80.69\% & 70.34\% \\
    YourTTS~\cite{Casanova22YourTTS}       & 96.55\% & 68.97\% & 79.31\% & 31.03\% & 68.97\% & 72.41\% & 93.10\% & 96.55\% & 79.31\% & 89.66\% & 83.45\% & 71.72\% \\
    Glow-TTS~\cite{Kim20GlowTTS}        & 100.00\% & 68.97\% & 68.97\% & 20.69\% & 72.41\% & 68.97\% & 82.76\% & 89.66\% & 72.41\% & 72.41\% & 79.31\% & 64.14\% \\
    TacoTron2-DDC~\cite{Shen18TacoTron2}   & 100.00\% & 68.97\% & 68.97\% & 17.24\% & 65.52\% & 75.86\% & 86.21\% & 96.55\% & 79.31\% & 72.41\% & 80.00\% & 66.21\% \\
    TacoTron2-DCA~\cite{Shen18TacoTron2, Gorodetskii22LongFormVC}   & 100.00\% & 65.52\% & 68.97\% & 13.79\% & 72.41\% & 72.41\% & 89.66\% & 96.55\% & 68.97\% & 72.41\% & 80.00\% & 64.14\% \\
    \bottomrule
  \end{tabular}
  }
\end{table*}



%% file: sections/evaluation.tex
\section{Evaluation}
\label{sec:evaluation}

\subsection{Experiment Setup}

\input{table/threshold}

\subsubsection{Experiment Settings}
We adopt five representative ASV models and six mainstream TTS models from SpeechBrain~\cite{Ravanelli21SpeechBrain, Ravanelli24SpeechBrainV1}, 3D-Speaker-Toolkit~\cite{Chen253DSpeakerToolkit} and Coqui-ai~\cite{CoquiTTS} as listed in Table~\ref{tab:asv_combined} and Appendix D. As our evaluation dataset, we select 100 utterances from a single speaker within the test-clean-100 subset of LibriSpeech~\cite{Panayotov15LibriSpeech}, ensuring each sample has a perfect alignment between the original recording and its TTS-cloned version across all selected TTS and ASV systems, more details can be find in Appendix C. This setup guarantees consistent voice identity across modalities. The resulting evaluation set contains audio samples ranging from 1.98 to 10.85 seconds, with an average length of 4.94 seconds.

\input{table/noise_levels}

In our \textit{Enkidu} framework, we attach UFP with a noise level fixed at 0.4. The UFP frame length is set to 120 frames, and the training ratio is set to 0.7, meaning 70\% of the samples are used to optimize the perturbation, while the remaining 30\% are reserved for evaluation. To ensure aligned simulation, we ensure the input texts for TTS generation exactly match those of the original audio. Experiment environment listed in Appendix F.

\subsubsection{Evaluation Metrics}
We evaluate privacy protection effectiveness from both shallow and deep perspectives, and assess the audio utility via both acoustic and intelligibility metrics.

\noindent\textbf{Shallow Protection Rate (SPR)}, as defined in Section~\ref{subsec:defense}, quantifies the extent to which an adversary's cloned audio fails to match the perturbed version under ASV verification. It reflects how well the perturbation prevents effective feature mimicry:
\begin{equation}
    \textbf{SPR} = \mathbb{E}_{x_i \sim \mathcal{D}_u}\left[ \mathbb{I}\left(\mathrm{SV}(\tilde{x}_i, \mathcal{G}(\tilde{x}_i, t);\tau) = 0 \right)\right],
\end{equation}

\noindent\textbf{Deep Protection Rate (DPR)}, also as refered in Section~\ref{subsec:defense}, measures whether the perturbation not only disrupts feature extraction but also misguides the entire synthesis process by TTS models:
\begin{equation}
    \textbf{DPR} = \mathbb{E}_{x_i \sim \mathcal{D}_u} \left[ \mathbb{I}\left(\mathrm{SV}(x_i, \mathcal{G}(\tilde{x}_i, t); \tau) = 0 \right) \right],
\end{equation}

\noindent\textbf{Threshold \& Equal Error Rate (EER)} are reported to ensure consistent and fair verification boundaries across different ASV systems. The system-specific thresholds are summarized in Table~\ref{tab:asv_combined}, and the computation methodology is detailed in Appendix B.

\noindent\textbf{Real-Time Coefficient (RTC)} quantifies efficiency as the ratio of processing time to input duration (in seconds). Lower RTC implies better real-time suitability.

\noindent\textbf{Mean Opinion Score (MOS) \& Short-Time Objective Intelligibility (STOI)} evaluate the perceptual quality of audio, respectively. MOS reflects subjective human listening experience and is estimated using MOSNet~\cite{Lo19MOSNet}, a non-intrusive, learning-based model. As a reference, the MOS of our custom dataset is 3.41 ± 0.28, indicating good baseline audio quality prior to perturbation. STOI~\cite{Taal10STOI} provides an objective, reference-based measure of quality, ranging from 0 to 1, with higher values indicating clearer speech. 


\noindent\textbf{Character Error Rate (CER)} \& \textbf{Word Error Rate (WER)} assess intelligibility via transcription accuracy—lower is better. We use Whisper~\cite{Radford23Whisper} to transcribe original and perturbed audio.

\input{table/efficiency}

\subsection{Performances}

\input{table/ablation_frame_length}

\input{table/ablation_train_ratio}

\subsubsection{Privacy-Preserving Performance Analysis}
Table~\ref{tab:main_experiment} shows that \textit{Enkidu} achieves strong privacy protection across diverse ASV and TTS models. The average Shallow Protection Rate SPR reaches 87.67\%, with particularly high scores on ECAPA-TDNN and ERes2Ne-t. DPR averages 68.12\%, and peaks at 100\% in several combinations. Performance is slightly lower on the X-Vector model, likely due to its relatively high EER, which weakens its discriminative power and reduces sensitivity to perturbation. Despite strong protection, Enkidu maintains excellent audio utility, with a MOS of 3.01 ± 0.07, a STOI of 0.71 ± 0.01 and perfect intelligibility (0.00\% CER and WER) as assessed by MOSNet and Whisper respectively.

\subsubsection{Real-Time Analysis}
We evaluate the runtime efficiency of \textit{Enkidu} using the RTC. As shown in Figure~\ref{subfig:rtc}, our method achieves a remarkably low RTC on GPU, consistently below 0.0006 across increasing audio lengths. This confirms the feasibility of real-time deployment, even on long-form audio. On CPU, the RTC remains under 0.001, though GPU acceleration yields better scalability.

\subsubsection{Processing Memory Analysis}
We analyze GPU memory consumption to assess scalability. As illustrated in Figure~\ref{subfig:mem}, both steady and peak memory usage increase linearly with audio length, reflecting predictable and controllable growth. Even for 60-second audio, the peak memory remains under 70MB, demonstrating that \textit{Enkidu} maintains extremely lightweight resource demands. Notably, the deployed \textit{Tiler} mentioned in Section~\ref{subsec:method} requires only 4MB, making it highly suitable for edge or embedded scenarios with strict memory constraints.

\subsection{Adaptive Attack Analysis}





\begin{figure}[]
  \centering
  \includegraphics[width=0.47\textwidth]{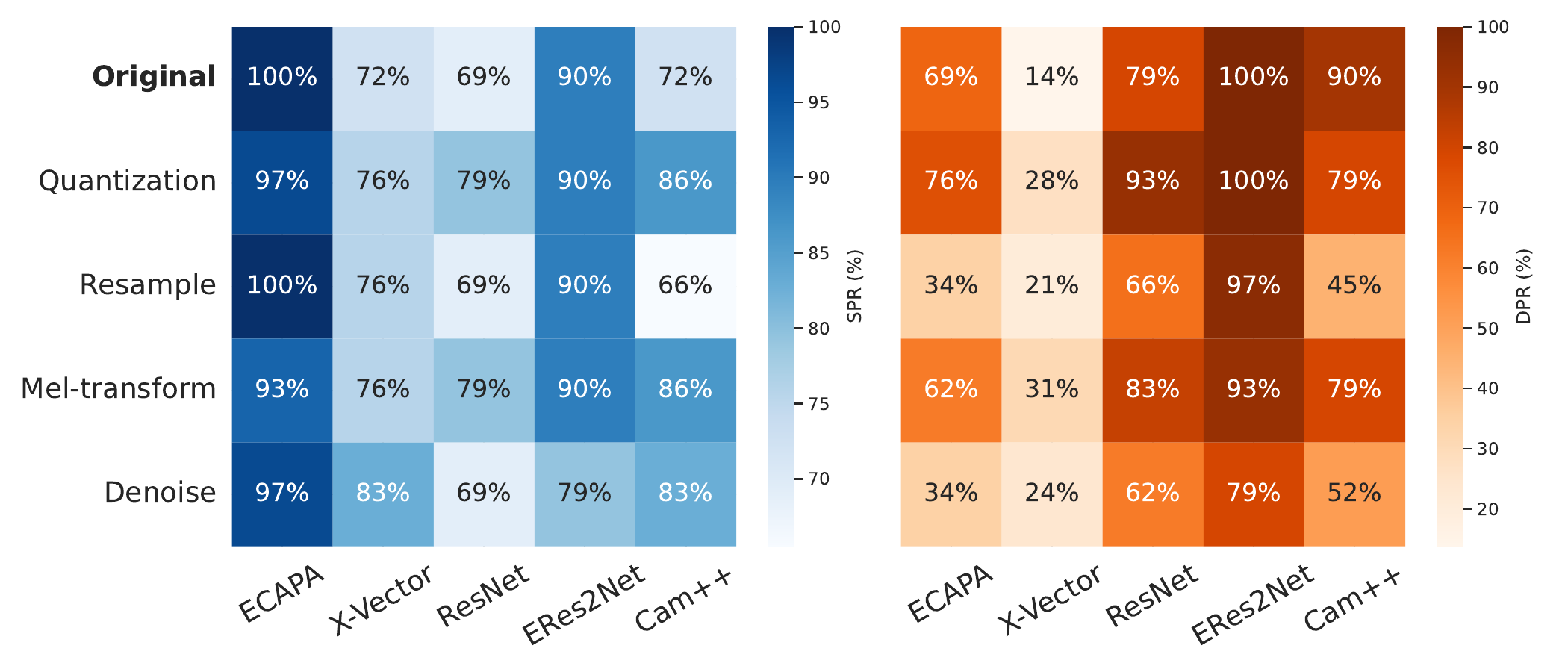}
  \caption{Adaptive attacker analysis. SPR and DPR heatmaps under four kinds of adaptive attack.}
  \label{fig:adaptive_attacker}
\end{figure}

As mentioned in Section~\ref{subsec:attacker}, an attacker may apply signal processing techniques to accessible audios in an attempt to suppress or remove the perturbation. To evaluate the robustness of \textit{Enkidu} against such adaptive threats, we consider four representative pre-processing attacks, as visualized in Figure~\ref{fig:adaptive_attacker} via heatmaps of SPR and DPR across five ASV models. Specifically, we examine: (1) \textbf{Quantization}, which reduces waveform precision by converting audio to 8-bit resolution and reconstructing it to approximate the original; (2) \textbf{Resample}, which downsamples the original 16kHz waveform to 8kHz and then upsamples it back to 16kHz; (3) \textbf{Mel-transform}, which converts the waveform into a mel-spectrogram and then reconstructs it via inverse transformation; and (4) \textbf{Denoise}, which applies Wiener filtering to suppress background noise and potential perturbation artifacts.


Notably, quantization and mel-transform are the most resilient cases, with average SPRs consistently above 85\%, and DPRs outperforming the original setting in some instances (e.g., ECAPA-TDNN's DPR rises from 68.97\% to 75.86\%). Denoising and resampling slightly reduce protection, especially on Cam++ and ResNet backbones, yet the DPR remains above 60\% in most settings.

These results demonstrate that \textit{Enkidu} is not only effective in standard settings but also exhibits notable robustness under adaptive attackers attempting to neutralize the perturbation.

\subsection{Ablation Analysis}

\subsubsection{Noise Level.}
We analyze the effect of varying the perturbation noise level on both privacy-preserving effectiveness and audio quality. As shown in Table~\ref{tab:noise_level}, increasing the noise level generally improves SPR and DPR across all ASV models, up to a point. The default setting of 0.4 achieves a balanced performance, with average SPR/DPR values remaining high while maintaining a MOS of 3.01 ± 0.07 and STOI of 0.71 ± 0.01. At low noise levels (e.g., 0.1–0.2), protection is weak due to insufficient perturbation energy. Conversely, higher noise levels ($\geq$ 0.6) offer slightly stronger SPR in some cases, but lead to degradation in acoustic quality—e.g., MOS drops to 2.39 at noise level 0.7, and STOI falls below 0.60. Importantly, intelligibility is preserved across all settings, with CER and WER consistently at 0.00\%. These results suggest that \textit{Enkidu} achieves effective protection even at modest noise levels, with 0.4 offering the best trade-off between utility and defense.

\subsubsection{Frame Length.}
We study the impact of varying the UFP frame length on privacy protection performance. As shown in Figure~\ref{fig:frame_length_ablation}, frame length significantly influences both SPR and DPR. A smaller frame length (e.g., 30) yields strong results, especially for DPR, reaching 100\% in most ASV models. However, performance tends to degrade with mid-range values (60–120), before recovering at larger lengths (240–300). This suggests that very short and very long frames offer better temporal alignment or frequency coverage, while mid-range values may introduce instability or over-smoothing in perturbation placement.

\subsubsection{Train Ratio.}
We also investigate the effect of training data scale by varying the train-test split ratio. As shown in Figure~\ref{fig:train_ratio_ablation}, protection performance improves steadily with more training data. Even with only 10\% training data, Enkidu achieves non-trivial privacy gains (e.g., ~36\% SPR on ECAPA-TDNN). Notably, SPR stabilizes near 100\% as the ratio approaches 0.7, while DPR continues to improve gradually. This confirms that \textit{Enkidu} is effective under few-shot conditions, and scales well with more user-provided samples.

\begin{figure}[t]
  \centering
  \includegraphics[width=0.47\textwidth]{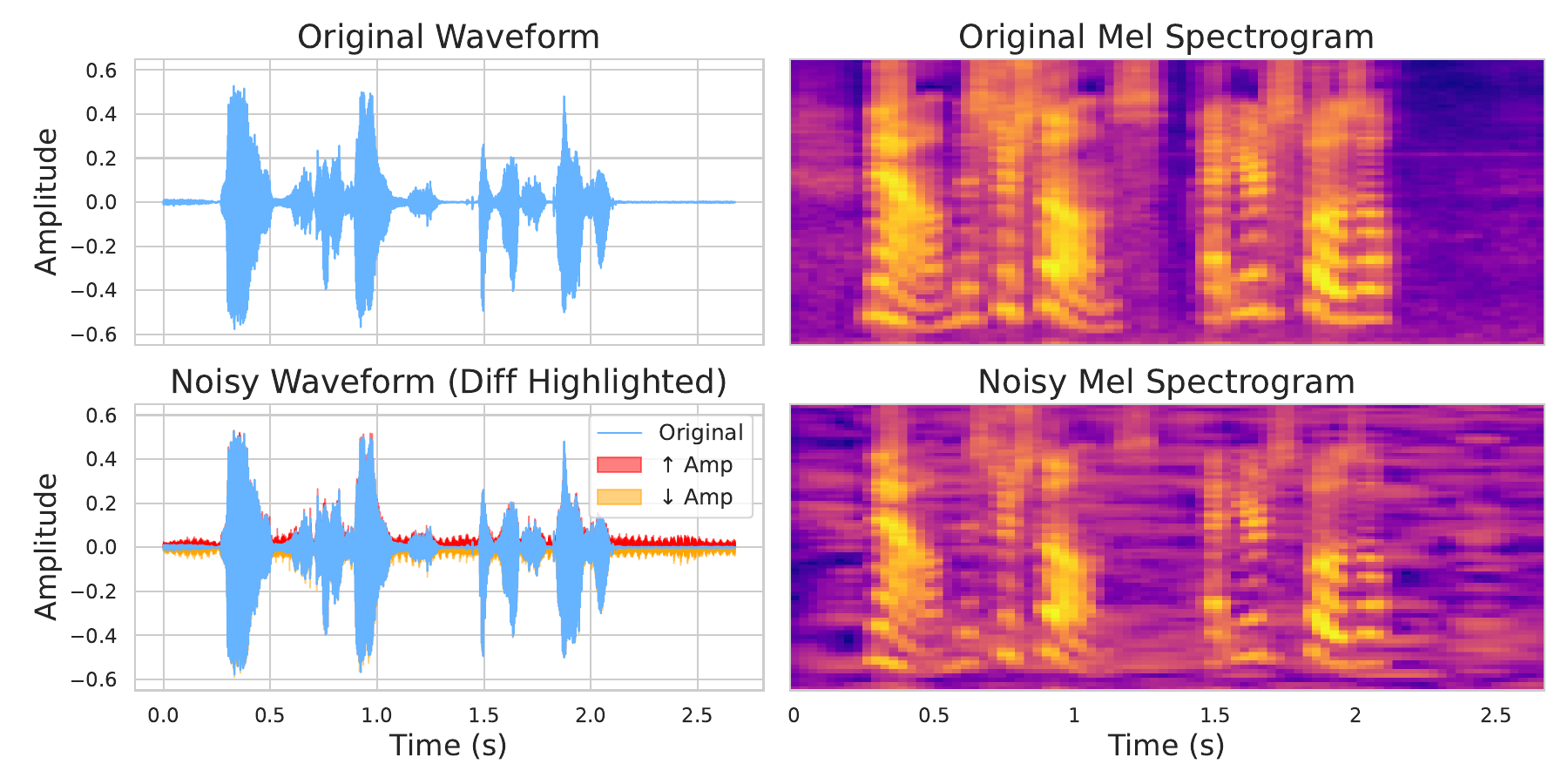}
  \caption{Time and frequency domain comparison between original and UFP-perturbed audio.}
  \label{fig:distortion_analysis}
\end{figure}

\subsection{Distortion Analysis}

To better understand how the UFP affects audio signals, we visualize both time-domain waveforms and mel spectrograms of original and perturbed audio, as shown in Figure~\ref{fig:distortion_analysis}. In the time domain, the waveform of the noisy sample resembles the original, with minor amplitude variations—indicating that the perturbation does not induce perceptible distortion to human ears.

In the frequency domain, the mel spectrogram of the noisy audio reveals subtle but structured frequency shifts. These perturbations are sufficient to confuse speaker verification models while remaining nearly imperceptible to human listeners, as corroborated by the MOS and intelligibility scores in prior sections. This visual evidence aligns with our psychoacoustic design principle: effective defense with minimal perceptual disturbance.

%% file: table/threshold.tex

\begin{table}[b]
  \centering
  \caption{ASV models info \& performance.}
  \label{tab:asv_combined}
  \resizebox{0.98\linewidth}{!}{
  \begin{tabular}{cccccc}
    \toprule
    \textbf{ASV Model} & \textbf{Train Set} & \textbf{Test Set} & \textbf{Params} & \textbf{EER} & \textbf{Threshold} \\
    \midrule
    ECAPA-TDNN~\cite{Desplanques20ECAPA} & VoxCeleb~\cite{Nagrani17VoxCeleb} & LibriSpeech & 14.66M & 1.20\% & 0.2922 \\
    X-Vector~\cite{Snyder18XVector} & VoxCeleb & LibriSpeech & 5.60M & 6.80\% & 0.9379 \\
    ResNet~\cite{He16ResNet, Zeinali19BUTVoxSRC} & VoxCeleb & LibriSpeech & 6.34M & 0.80\% & 0.3136 \\
    Cam++~\cite{Wang23CAMpp} & VoxCeleb & LibriSpeech & 7.18M & 1.00\% & 0.3770 \\
    ERes2Net~\cite{Chen23EnhancedRes2Net} & VoxCeleb & LibriSpeech & 22.46M & 0.80\% & 0.3950 \\
    \bottomrule
  \end{tabular}
  }
\end{table}

%% file: table/noise_levels.tex
\begin{table*}[t]
  \caption{
    Impact of different noise levels on protection effectiveness and audio quality. Higher SPR/DPR indicate stronger defense performance. Enkidu maintains intelligibility with CER and WER at 0.00\%±0.00\% across all noise levels.
  }
  \label{tab:noise_level}
  \centering
  \resizebox{0.98\linewidth}{!}{
  \begin{tabular}{ccccccccccccccccc}
    \toprule
    \multirow{2}{*}{\textbf{Noise Level}} & \multicolumn{2}{c}{\textbf{ECAPA-TDNN}} & \multicolumn{2}{c}{\textbf{X-Vector}} & \multicolumn{2}{c}{\textbf{ResNet}} & \multicolumn{2}{c}{\textbf{ERes2Net}} & \multicolumn{2}{c}{\textbf{Cam++}} & \multirow{2}{*}{\textbf{MOS}} & \multirow{2}{*}{\textbf{STOI}} & \multirow{2}{*}{\textbf{CER}} & \multirow{2}{*}{\textbf{WER}} \\
    \cline{2-11}
    & \textbf{SPR} & \textbf{DPR} & \textbf{SPR} & \textbf{DPR} & \textbf{SPR} & \textbf{DPR} & \textbf{SPR} & \textbf{DPR} & \textbf{SPR} & \textbf{DPR} & & & & \\
    \midrule
    0.1 & 10.34\% & 10.34\% & 13.79\% & 13.79\% & 24.14\% & 24.14\% & 34.48\% & 41.38\% & 24.14\% & 24.14\% & 3.33±0.24 & 0.98±0.00 & 0.00\%±0.00\% & 0.00\%±0.00\% \\
    0.2 & 72.41\% & 27.59\% & 55.17\% & 20.69\% & 41.38\% & 51.72\% & 51.72\% & 72.41\% & 58.62\% & 48.28\% & 2.97±0.09 & 0.86±0.00 & 0.00\%±0.00\% & 0.00\%±0.00\% \\
    0.3 & 89.66\% & 37.93\% & 62.07\% & 13.79\% & 55.17\% & 58.62\% & 65.52\% & 86.21\% & 79.31\% & 58.62\% & 2.88±0.10 & 0.80±0.00 & 0.00\%±0.00\% & 0.00\%±0.00\% \\
    0.4 (Default) & 100.00\% & 68.97\% & 72.41\% & 13.79\% & 68.97\% & 79.31\% & 89.66\% & 100.00\% & 72.41\% & 89.66\% & 3.01±0.07 & 0.71±0.01 & 0.00\%±0.00\% & 0.00\%±0.00\% \\
    0.5 & 96.55\% & 62.07\% & 75.86\% & 24.14\% & 75.86\% & 89.66\% & 96.55\% & 96.55\% & 82.76\% & 86.21\% & 2.84±0.06 & 0.68±0.01 & 0.00\%±0.00\% & 0.00\%±0.00\% \\
    0.6 & 93.10\% & 34.48\% & 79.31\% & 17.24\% & 68.97\% & 68.97\% & 82.76\% & 68.97\% & 65.52\% & 51.72\% & 2.62±0.06 & 0.63±0.01 & 0.00\%±0.00\% & 0.00\%±0.00\% \\
    0.7 & 96.55\% & 62.07\% & 86.21\% & 24.14\% & 89.66\% & 79.31\% & 100.00\% & 96.55\% & 86.21\% & 68.97\% & 2.39±0.08 & 0.60±0.01 & 0.00\%±0.00\% & 0.00\%±0.00\% \\
    0.8 & 96.55\% & 58.62\% & 82.76\% & 24.14\% & 79.31\% & 79.31\% & 93.10\% & 96.55\% & 72.41\% & 65.52\% & 2.73±0.07 & 0.60±0.01 & 0.00\%±0.00\% & 0.00\%±0.00\% \\
    0.9 & 96.55\% & 51.72\% & 93.10\% & 27.59\% & 96.55\% & 89.66\% & 96.55\% & 96.55\% & 79.31\% & 58.62\% & 2.57±0.07 & 0.57±0.01 & 0.00\%±0.00\% & 0.00\%±0.00\% \\
    1.0 & 93.10\% & 55.17\% & 89.66\% & 27.59\% & 89.66\% & 79.31\% & 93.10\% & 89.66\% & 82.76\% & 62.07\% & 2.62±0.05 & 0.57±0.01 & 0.00\%±0.00\% & 0.00\%±0.00\% \\
    \bottomrule
  \end{tabular}
  }
\end{table*}

%% file: table/efficiency.tex

\begin{figure}[]
    \centering
    \begin{subfigure}[b]{0.8\linewidth}
        \centering
        \includegraphics[width=\linewidth]{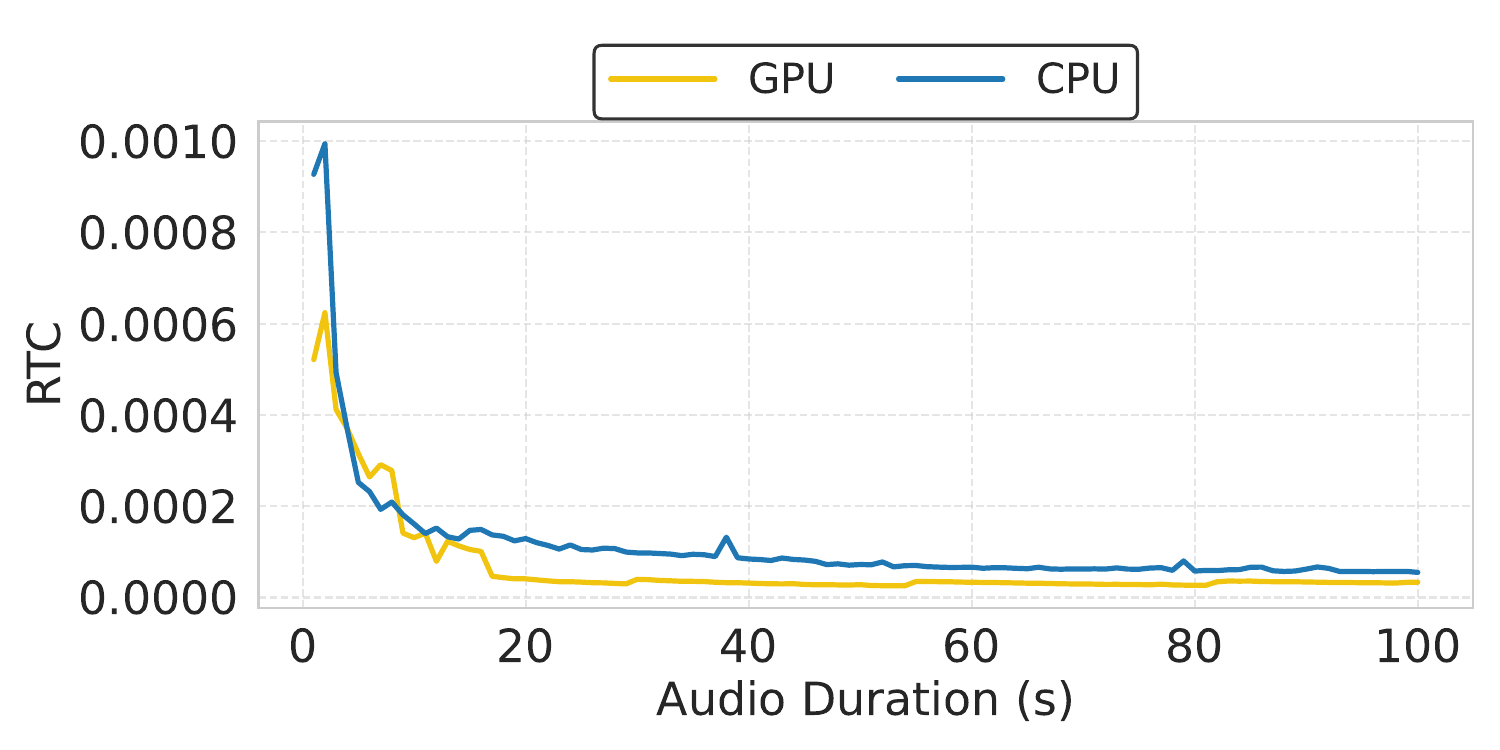}
        \caption{Real-time analysis}
        \label{subfig:rtc}
    \end{subfigure}
    \vskip\baselineskip
    \begin{subfigure}[b]{0.8\linewidth}
        \centering
        \includegraphics[width=\linewidth]{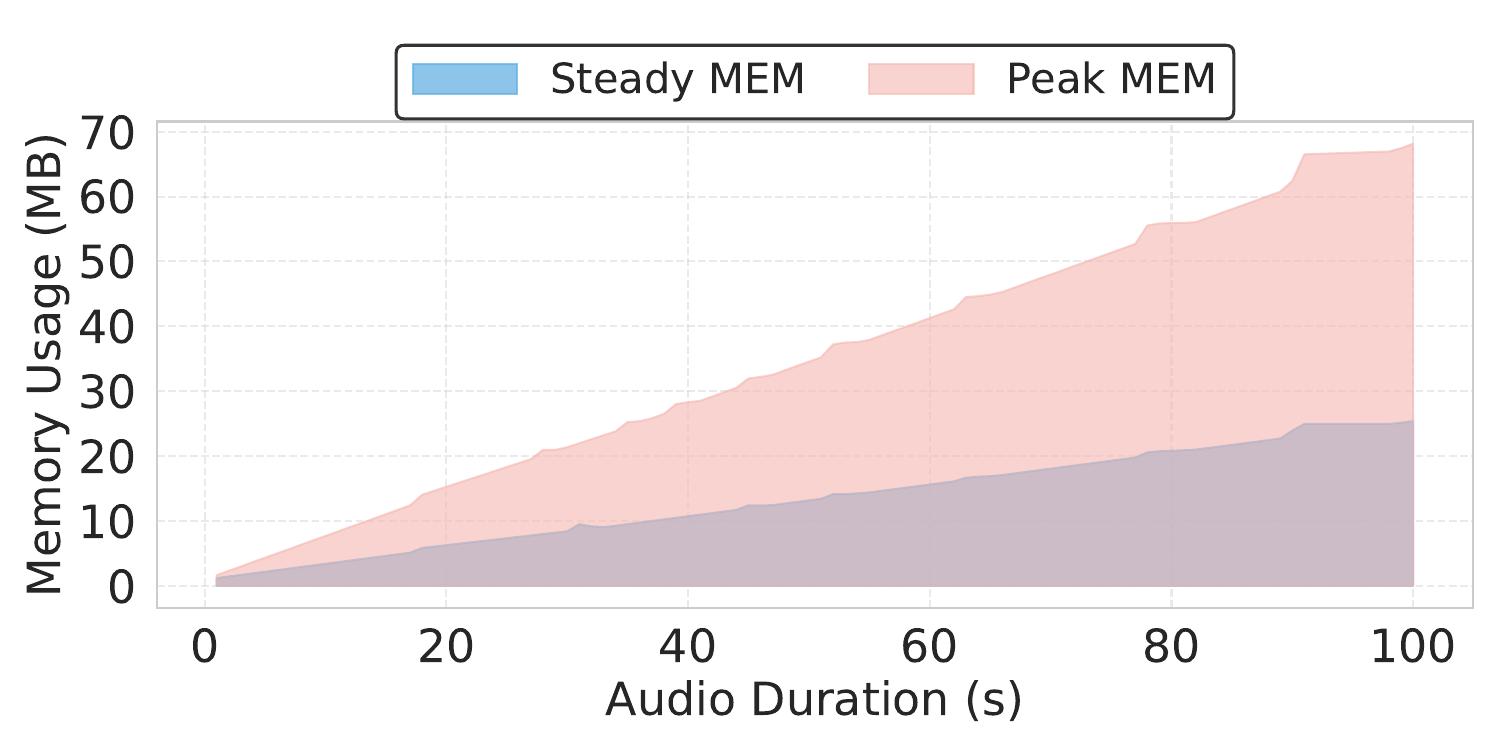}
        \caption{Processing memory analysis}
        \label{subfig:mem}
    \end{subfigure}
    \caption{UFP efficiency analysis. Across varying audio durations (1–100 seconds at 16kHz).}
    \label{fig:efficiency}
\end{figure}

%% file: table/ablation_frame_length.tex

\begin{figure*}[]
    \centering
    \begin{subfigure}[b]{0.18\textwidth}
        \centering
        \includegraphics[width=\textwidth]{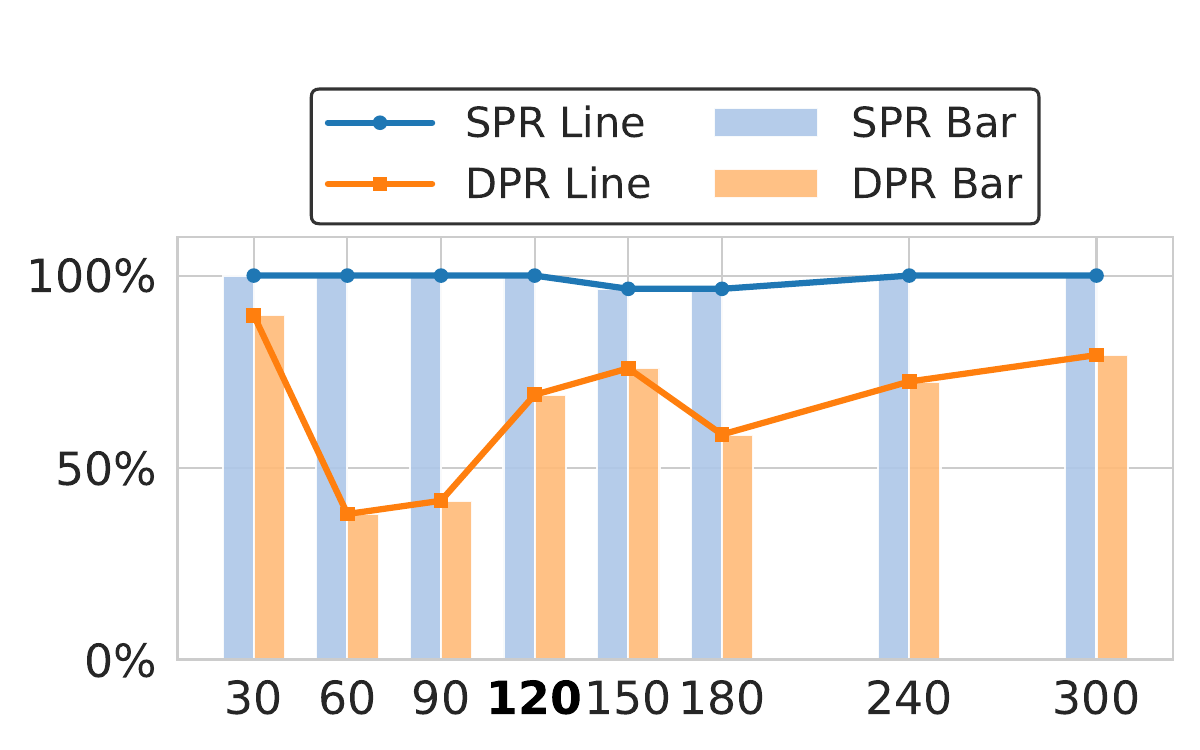}
        \caption{ECAPA-TDNN}
    \end{subfigure}
    \begin{subfigure}[b]{0.18\textwidth}
        \centering
        \includegraphics[width=\textwidth]{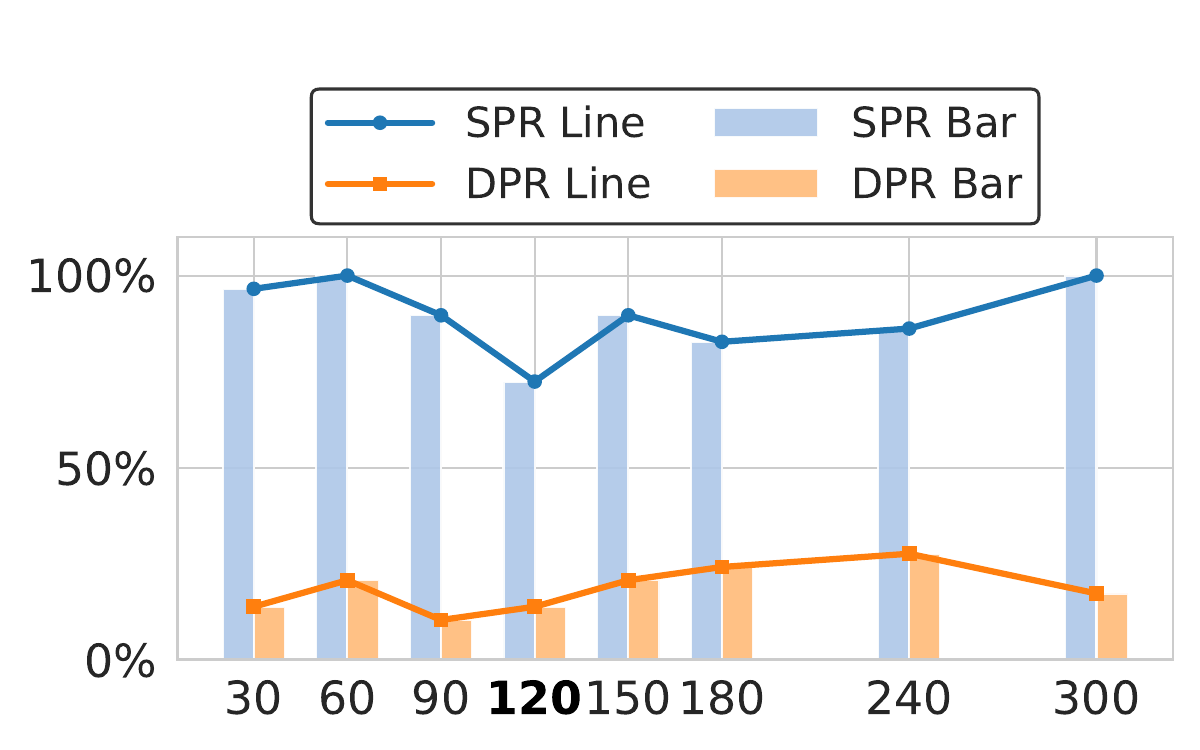}
        \caption{X-Vector}
    \end{subfigure}
    \begin{subfigure}[b]{0.18\textwidth}
        \centering
        \includegraphics[width=\textwidth]{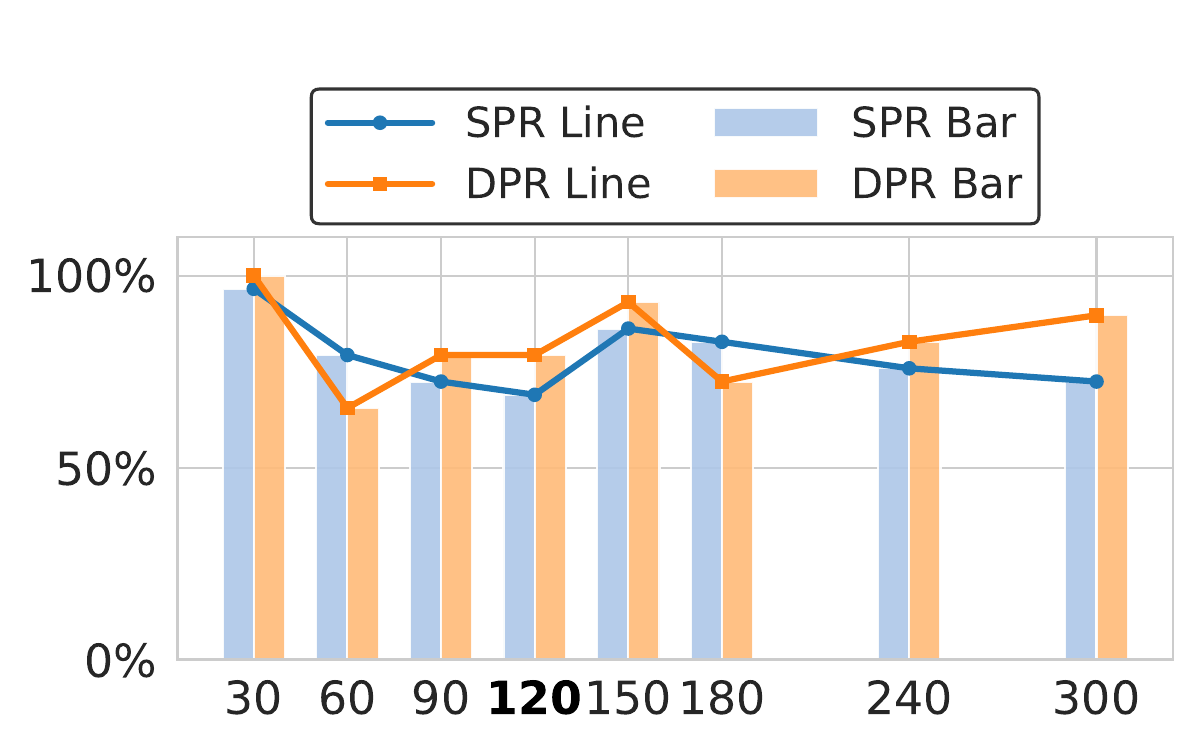}
        \caption{ResNet}
    \end{subfigure}
    \begin{subfigure}[b]{0.18\textwidth}
        \centering
        \includegraphics[width=\textwidth]{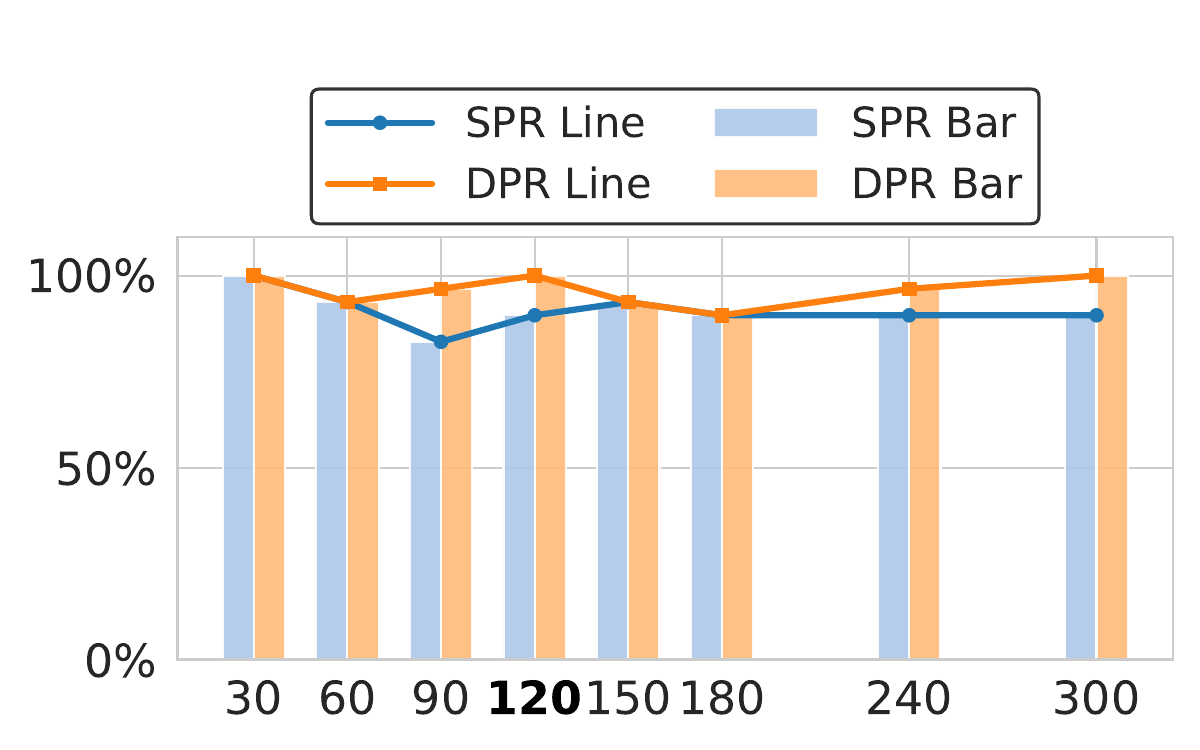}
        \caption{ERes2Net}
    \end{subfigure}
    \begin{subfigure}[b]{0.18\textwidth}
        \centering
        \includegraphics[width=\textwidth]{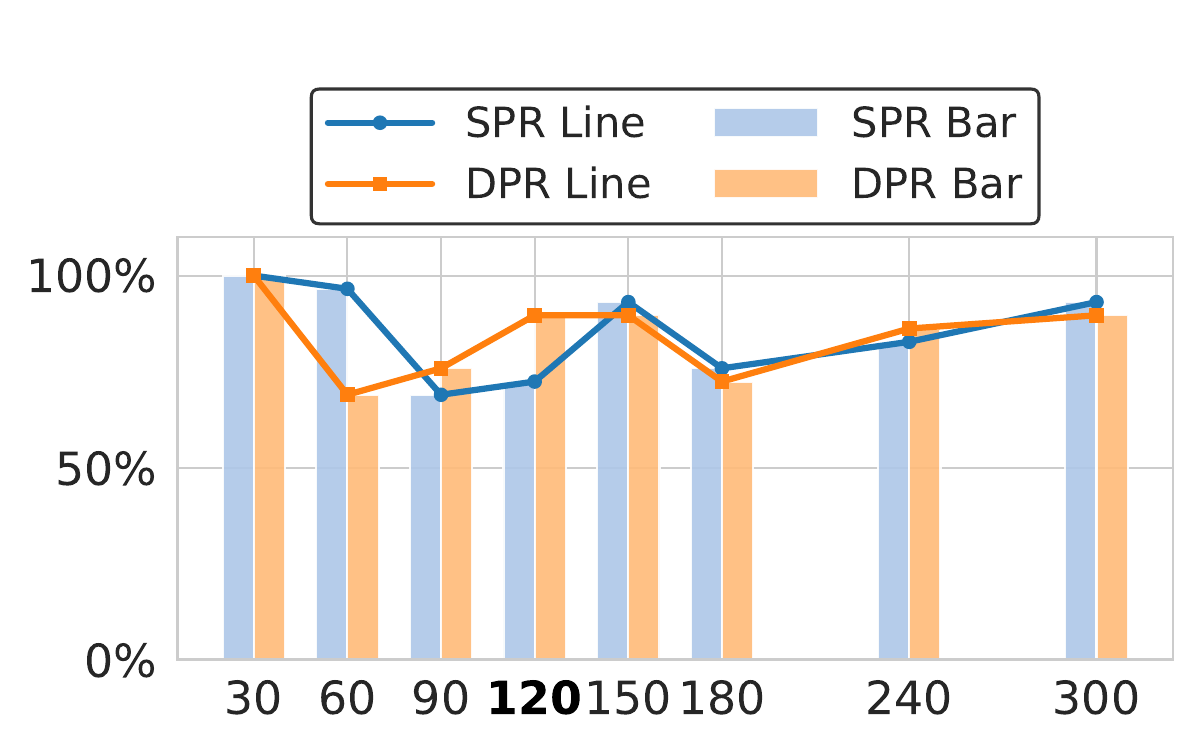}
        \caption{Cam++}
    \end{subfigure}
    \caption{Ablation results under different \textbf{Frame Length} settings across ASV models. Both SPR (bar) and DPR (line) are visualized to highlight trade-offs in temporal perturbation granularity.}
    \label{fig:frame_length_ablation}
\end{figure*}

%% file: table/ablation_train_ratio.tex

\begin{figure*}[]
    \centering
    \begin{subfigure}[b]{0.18\textwidth}
        \centering
        \includegraphics[width=\textwidth]{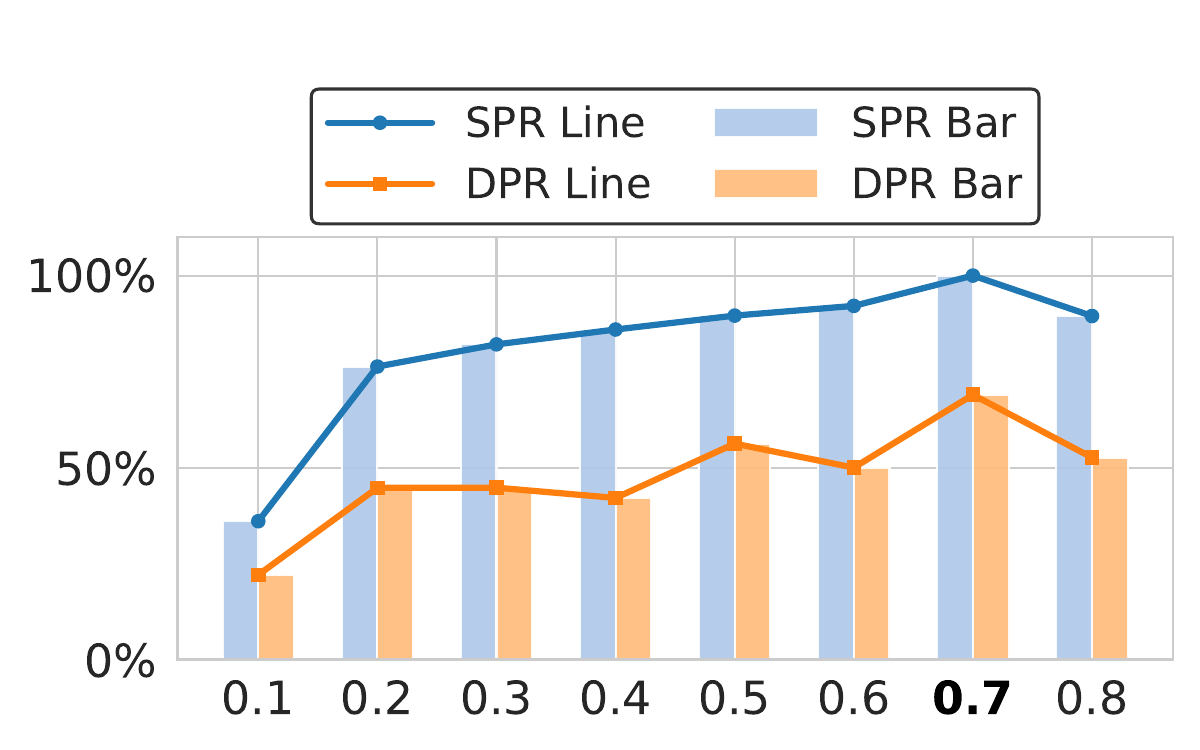}
        \caption{ECAPA-TDNN}
    \end{subfigure}
    \begin{subfigure}[b]{0.18\textwidth}
        \centering
        \includegraphics[width=\textwidth]{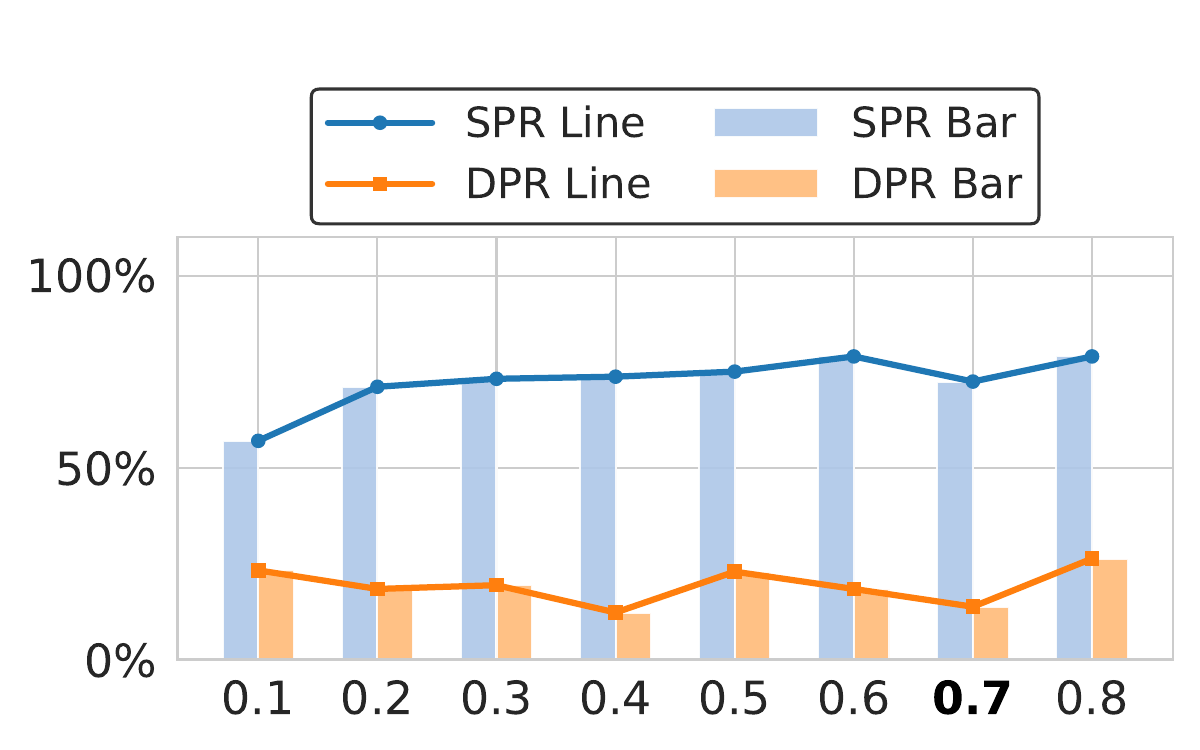}
        \caption{X-Vector}
    \end{subfigure}
    \begin{subfigure}[b]{0.18\textwidth}
        \centering
        \includegraphics[width=\textwidth]{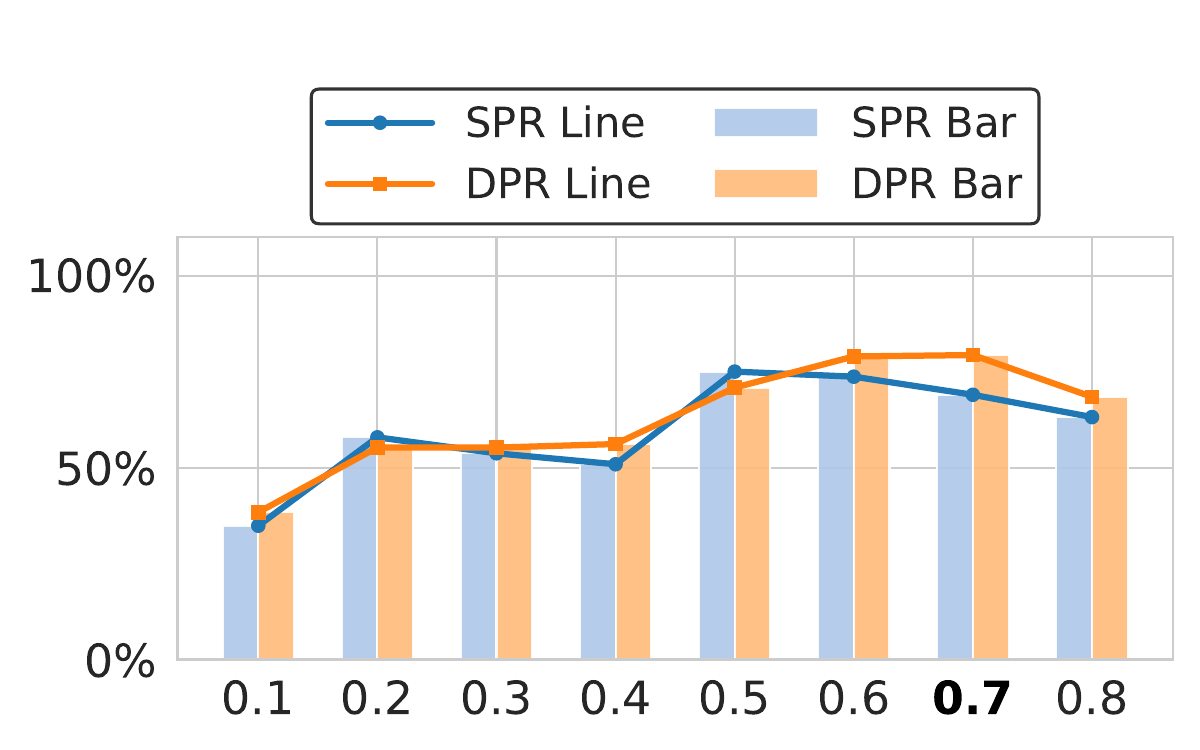}
        \caption{ResNet}
    \end{subfigure}
    \begin{subfigure}[b]{0.18\textwidth}
        \centering
        \includegraphics[width=\textwidth]{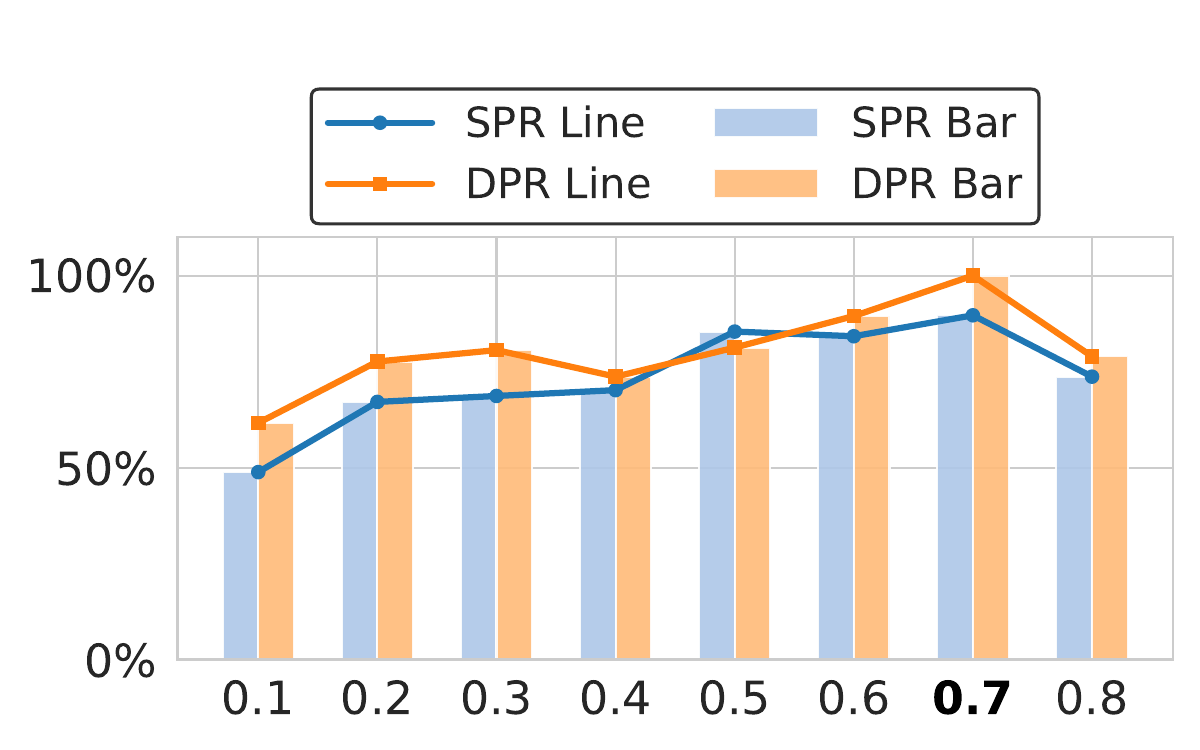}
        \caption{ERes2Net}
    \end{subfigure}
    \begin{subfigure}[b]{0.18\textwidth}
        \centering
        \includegraphics[width=\textwidth]{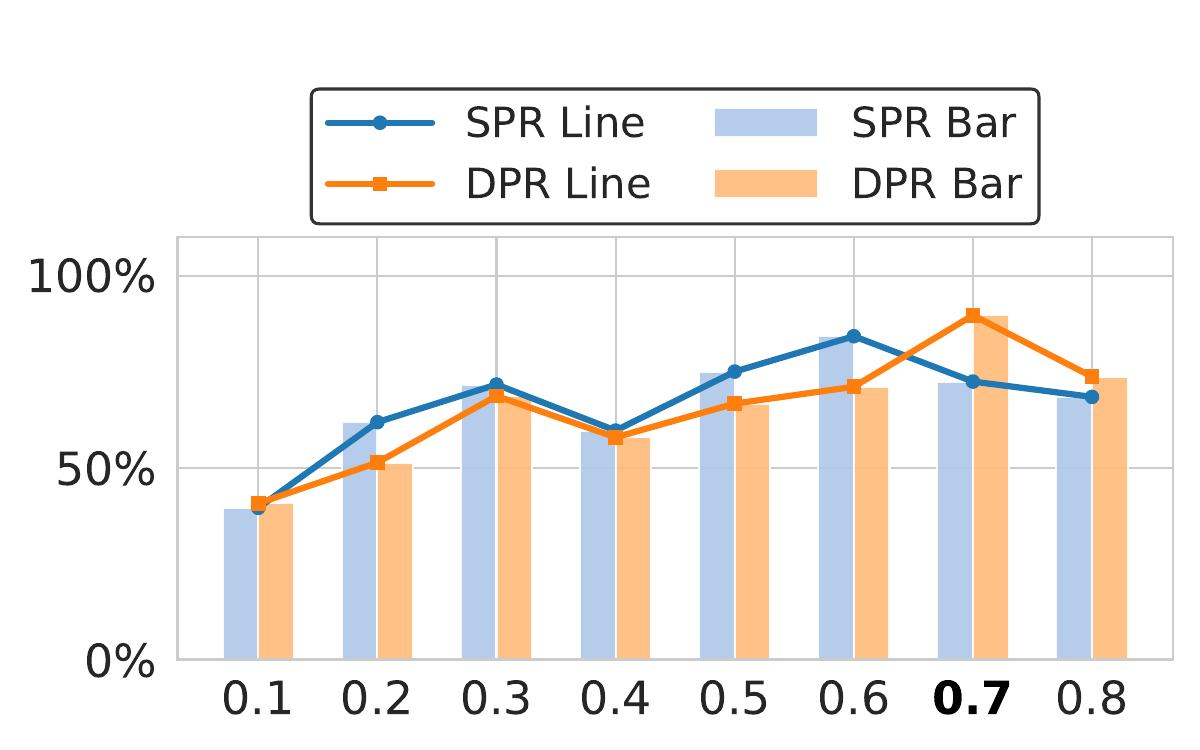}
        \caption{Cam++}
    \end{subfigure}
    \caption{Ablation results under different \textbf{Train Ratios}. Even with limited training data, \textit{Enkidu} achieves strong SPR/DPR, and performance scales consistently with increased data availability.}
    \label{fig:train_ratio_ablation}
\end{figure*}

%% file: sections/conclusion.tex


\section{Conclusion}


In this work, we present \textit{Enkidu}, a universal and user-oriented framework for real-time audio privacy protection against voice-based deepfake threats. By leveraging few-shot optimization, \textit{Enkidu} generates a UFP that preserves audio quality while significantly reducing speaker similarity in black-box settings. Our method demonstrates strong transferability across unseen user audio utterances, supports arbitrary-length audio, and operates with low computational and temporal costs, making it well-suited for practical deployment in resource-constrained environments. Extensive evaluations against SOTA TTS systems validate its effectiveness in mitigating deepfake risks while maintaining real-time applicability.

%% file: main.bbl

%% file: sections/appendix.tex
\appendix

\section{Voice Data}
Voice data possesses unique temporal, spectral, and prosodic characteristics, distinguishing it significantly from other multimedia data forms. Given its sensitivity and frequent inclusion of personal attributes, the internet giants more os less have proposed relevant policies for voice data privacy-preservation. 

\input{table/tts_match_rate}

Structurally, voice data comprises two main components:
\begin{itemize}
    \item \textit{Voice Features}: These include timbral qualities, pitch, intonation patterns, and vocal tract characteristics. Such features are fundamental for speaker identification tasks and recent speech synthesis process.
    \item \textit{Speech Content}: This refers to the semantic and contextual information embedded in audio data. 
\end{itemize}

Effective utilization of voice data requires meticulous preprocessing. Typical preprocessing steps include noise reduction, normalization of audio amplitude, and silence trimming, all of which enhance data consistency. Subsequently, processed audio is converted into standardized acoustic representations, commonly Mel-spectrograms via Mel-filter bank, capturing essential acoustic information while reducing dimensional complexity. Augmentation techniques, such as adding controlled noise or altering pitch and temporal features, further enhance data robustness and model generalization capabilities.

\input{table/tts_tables}

\section{Threshold Setting for Speaker Verification}
\label{appendix:threshold}

To determine whether two audio samples originate from the same speaker, we adopt a cosine similarity-based verification strategy. Given two speaker embeddings $\mathbf{e}_1$ and $\mathbf{e}_2$ extracted from audio samples using a pretrained automatic speaker verification (ASV) model, we compute their similarity as:

\begin{equation}
\text{Sim}(\mathbf{e}_1, \mathbf{e}_2) = \frac{\mathbf{e}_1 \cdot \mathbf{e}_2}{\|\mathbf{e}_1\|_2 \cdot \|\mathbf{e}_2\|_2}.
\end{equation}

To establish a binary decision threshold, we generate a list of evaluation trials $\mathcal{T} = \{ (\mathbf{e}_i, \mathbf{e}_j, y_{ij}) \}$, where $y_{ij} \in \{0, 1\}$ indicates whether the pair is from the same speaker ($y_{ij}=1$) or not ($y_{ij}=0$). For each trial, we compute the similarity score and store it alongside the ground-truth label.

We then compute the false negative rate (FNR) and false positive rate (FPR) over all sorted similarity scores using cumulative sums:

\begin{align}
\text{FNR}(k) &= \frac{\sum_{i=1}^{k} w_i \cdot \mathbb{I}[y_i = 1]}{\sum_{i=1}^{N} w_i \cdot \mathbb{I}[y_i = 1]}, \\
\text{FPR}(k) &= 1 - \frac{\sum_{i=1}^{k} w_i \cdot \mathbb{I}[y_i = 0]}{\sum_{i=1}^{N} w_i \cdot \mathbb{I}[y_i = 0]},
\end{align}

where $w_i$ denotes the optional trial weight (uniform by default). From the FNR and FPR curves, we identify the Equal Error Rate (EER) point where $\text{FNR} = \text{FPR}$ and extract the corresponding similarity score as the decision threshold $\tau$:

\begin{equation}
\tau = \arg\min_{s} |\text{FNR}(s) - \text{FPR}(s)|.
\end{equation}

In our experiments, this process yields thresholds for five ASV systems, which we adopt throughout the evaluations. This data-driven approach ensures that the verification decision is calibrated according to real distributions of speaker similarity, and it provides a robust trade-off between false acceptance and false rejection under various conditions.

\begin{algorithm}
\caption{\textit{Tiler} Design}
\label{alg:Tiler}
\begin{algorithmic}
\Require Audio Sample $x$; Real Part of UFP $\delta_r$; Imaginary Part of UFP $\delta_i$; Noise level $\eta$; Frame Length $L_u$; Augmentation Bool $a$ 
\Ensure Perturbed Sample $\tilde{x}$

\Function{Tiler}{$X, \delta_r, \delta_i, \eta, L_u, a$}
    \State // Smoothing the frequential disturbance
    \State $\delta \gets \textsc{FreqSmoother}(\delta_{r}, \delta_{i})$
    \State // Short-Time Fourier Transform
    \State $S \gets \text{STFT}(x)$
    \State $n \gets \lfloor |S| / L_u \rfloor$
    \If{$a = 1$}
        \State // During the UFP optimization
        \State $r \gets \text{Mask Ratio}$, $\epsilon \gets \text{RandInt}(0,L_u)$
        \State $m \sim \text{Bernoulli}(1 - r)^n$
    \Else
        \State // In Deployment
        \State $m \gets \mathbf{1}^n$, $\epsilon \gets 0$
    \EndIf
    \For{$i = 0$ to $n - 1$}
        \If{$m_i = 1$}
            \State $S[:, \epsilon + i f : \epsilon + (i + 1) f] \;\; $+=$ \;\; \eta \cdot \delta$
        \EndIf
    \EndFor
    \State // Inverse Transform to Time Domain
    \State $\tilde{x} \gets \text{iSTFT}(S)$
    \State \Return $\tilde{x}$
\EndFunction

\Function{FreqSmoother}{$\delta_r, \delta_i$, $k=5$}
    \State $K \gets \frac{1}{k} \cdot \mathbf{1}_{1 \times k}$
    \State $\delta_r^{-} \gets \text{Conv1D}(\delta_r, K, \text{pad} = \lfloor k/2 \rfloor)$
    \State $\delta_i^{-} \gets \text{Conv1D}(\delta_i, K, \text{pad} = \lfloor k/2 \rfloor)$
    \State $\delta \gets \delta_r^{-} + j \cdot \delta_i^{-}$
    \State \Return $\delta$
\EndFunction
\end{algorithmic}
\end{algorithm}

\section{Deepfake Audios Analysis}
\label{appendix:match_rate}

To better understand the vulnerability of raw audio to voice deepfake attacks, we evaluate the \textbf{match rate} between deepfake audio and its source speaker's embedding across five distinct ASV backends.

As shown in Table~\ref{tab:match_rate_unprotected}, we test six TTS systems—SpeedySpeech, FastPitch, YourTTS, Glow-TTS, Tacotron2-DDC, and Tacotron2-DCA, as same as the settings in main body of paper, against ECAPA-TDNN, X-Vector, ResNet, ERes2Net, and Cam++ ASV models.

The \textbf{match rate} is defined as the percentage of fake audio samples that pass the ASV verification when compared with their source utterances, using the threshold $\tau$ defined in Appendix~\ref{appendix:threshold}. A higher match rate indicates more successful mimicry of the speaker's identity.

\paragraph{Findings.} 
Across all ASV systems, the X-Vector consistently exhibits the highest match rate (above 92\% across all TTS models), revealing its high susceptibility to TTS-based impersonation. Conversely, ERes2Net appears more robust, with match rates often below 52\%. Among the TTS models, TacoTron2-DCA achieves the highest average match rate (72.0\%), indicating its strong ability to synthesize speaker-indistinguishable audio. These results highlight the critical privacy threat posed by modern TTS systems when users’ raw audio remains unprotected.

This analysis reinforces the necessity for a universal and real-time audio protection mechanism, as proposed in our main method.

\section{Model Info}
\label{appendix:model_info}

The specifications and architectural details of the TTS models used in our evaluation are provided in Table~\ref{tab:tts_table}. These models represent a diverse set of modern voice synthesis techniques.

\section{\textit{Tiler} Algorithm Details}
\label{appendix:tiler}

The procedural implementation of the proposed \textit{Tiler} algorithm is presented in Algorithm~\ref{alg:Tiler}. This algorithm plays a central role in the construction and application of the UFP described in the main body of the paper.

\section{Experiment Environment}

All experiments are conducted on a high-performance server equipped with Intel(R) Xeon(R) Platinum 8358P CPUs (3.40GHz), 386GB RAM, and an NVIDIA A800 GPU. The implementation environment is based on VSCode and PyTorch.

\section{Theoretical Analysis}
\label{subsec:freq_advantage}

\subsection{Preliminaries and Assumptions}

Let $x\!\in\!\mathbb R^{T}$ be a speech waveform and
$S=\operatorname{STFT}(x)\!\in\!\mathbb C^{1\times B\times L}$ its complex spectrogram,
where $B$ is the number of frequency bins and $L$ the number of frames
($T\!=\!L H$, hop size $H$).
A universal frequential perturbation (UFP) is a tensor
$\delta\!=\!\delta_r+j\delta_i\in\mathbb C^{1\times B\times L_u}$ with $L_u\!\ll\!L$.
In both optimisation and deployment, \textit{Tiler} applies
\begin{equation}
   \tilde S[:,m]
     = S[:,m] + \delta[:,\,m\bmod L_u],\quad 0\le m<L,
   \label{eq:tiling}
\end{equation}
and returns $\tilde x=\operatorname{iSTFT}(\tilde S)$.
We adopt two standard assumptions:

\begin{enumerate}
    \item \textbf{Short-term stationarity}: Within a window of $L_u$ frames,
          the acoustic statistics of speech are approximately constant.%
          \label{ass:stationarity}
    \item \textbf{Energy orthogonality}: The STFT uses an analysis window
          $w[n]$ satisfying $\sum_{m} w[n-mH]\,w[n'-mH]\!=\!0$ for $n\!\neq\!n'$
          (e.g.\ Hann), ensuring per-frame energy additivity.%
          \label{ass:orthogonality}
\end{enumerate}

\subsection{Problem Re-statement}

\input{table/comparsion_data}

With Equation~\ref{eq:opt_goal}, the
defender solves\footnote{%
 The perceptual-quality term is kept identical in both domains to isolate the
 effect of the optimisation space.}
\begin{equation}
   \min_{\delta}\;
   \mathbb E_{x_i\sim\mathcal D_u}
        \Big[L_{\text{fea}}\bigl(\mathcal E(x_i),
              \mathcal E(\tilde x_i)\bigr)\Big]
   \quad\text{s.t.}\quad
   \|\delta\|_{p}\le\varepsilon.
   \label{eq:freq_obj}
\end{equation}
If we worked in the time domain, we would instead learn
$v\!\in\!\mathbb R^{T}$ subject to $\|v\|_p\!\le\!\varepsilon$ and set
$\tilde x = x+v$.

\subsection{Main Results}

\paragraph{Proposition 1 (Parameter-Efficiency).}
Under Assumption~\ref{ass:stationarity}, the optimal frequency-domain
solution requires at most
\[
   P_{\text{freq}} = 2\,B\,L_u
   \quad\text{s.t.}\quad
   L_u \approx \tfrac{H}{w_s}\,,
\]
where $w_s$ is the stationarity window (typically $L_u\!\approx\!80\!-\!120$).
A time-domain universal perturbation of the same coverage length $T$ needs
\[
   P_{\text{time}} = T \approx \tfrac{L\,H}{w_s}\,,
\]
so $P_{\text{freq}}/P_{\text{time}}\!\approx\!2B/L\!\ll\!1$ (e.g.\ $<\!0.04$ for
$B\!=\!513$, $L\!=\!4000$).
\textit{Fewer degrees of freedom make optimisation faster and less prone to
over-fitting the small user set~$\mathcal D_u$.}

\textit{Sketch proof}.  Dimensionality counts follow directly from the shape
of $\delta$ in Equation~\ref{eq:tiling} and the linear relation $T\!=\!LH$.

\paragraph{Proposition 2 (Gradient Amplification).}
Let $\mathcal L_{\text{fea}}$ in Equation~\ref{eq:freq_obj} be differentiable.
Then
\begin{equation}
  \nabla_{\delta}\,
  \mathbb E_{x_i}\!
  \bigl[\mathcal L_{\text{fea}}\bigr]
  = \sum_{m=0}^{L-1}
    \mathbb E_{x_i}\!
    \left[\frac{\partial\mathcal L_{\text{fea}}}
               {\partial|S_i[:,m]|}\right].
  \label{eq:grad-amp}
\end{equation}
Because the same $\delta$ affects \emph{all} $L$ frames
(cf.\ Equation~\ref{eq:tiling}),
each SGD step aggregates $L$ local gradients, whereas a time-domain perturbation
updates one sample per location.
\textit{Thus the expected gradient norm in the frequency domain is
$\Theta(L)$ times larger, accelerating convergence.}

\textit{Sketch proof}.  Differentiate Equation~\ref{eq:tiling}, use linearity of
STFT and chain rule.

\paragraph{Proposition 3 (Shift-Equivariance).}
Let $x^\tau[n]=x[n-\tau]$ be a temporal shift.
Under Assumption~\ref{ass:orthogonality},
\[
   \operatorname{STFT}(x^\tau)[:,m]
      = S[:,m-\tau/H].
\]
Applying Equation~\ref{eq:tiling} then yields
$\tilde S^\tau[:,m]=\tilde S[:,m-\tau/H]$,
and hence $\tilde x^\tau[n]=\tilde x[n-\tau]$ after iSTFT.
Therefore the protection effect is invariant to arbitrary
\(\tau\), making the perturbation robust to latency, cropping, or padding.

\paragraph{Proposition 4 (Masking-Constraint Simplicity).}
Let $T[b]$ be the psychoacoustic masking threshold (\si{dB}) for bin $b$.
The feasible set
\(
   \mathcal C_{\text{freq}}
     :=\{\delta:|\delta[b]|<T[b],\;1\!\le\!b\!\le\!B\}
\)
is a \emph{convex box} in $\mathbb R^{2B L_u}$.
The corresponding time-domain constraint—
“instantaneous SPL below critical-band mask”—
is non-convex and couples all $T$ samples via a quadratic STFT operator.
Hence projections onto $\mathcal C_{\text{freq}}$
(cost $\mathcal O(B L_u)$) are \textbf{analytically and computationally cheaper}.

\section{Supplement Experiments}

\subsection{Experimental Setup and Datasets}
To rigorously benchmark \textit{Enkidu} against SOTA audio privacy methods, we conduct comprehensive experiments across three major speech datasets:
\begin{itemize}
    \item \textbf{LibriSpeech (English):} A large-scale corpus of read English speech, widely adopted in speaker recognition and TTS evaluation.
    \item \textbf{CommonVoice (French)~\cite{Ardila20CommonVoice}:} A multilingual open-source corpus, here focusing on French utterances for cross-lingual robustness.
    \item \textbf{AISHELL (Chinese)~\cite{Bu17AISHELL}:} A Mandarin speech dataset to assess performance on tonal and non-English languages.
\end{itemize}
Each dataset is evaluated with five prominent ASV backends: ECAPA-TDNN, X-Vector, ResNet, ERes2Net, and Cam++. For intelligibility and quality, we report CER, WER, MOS, and STOI as the Section~\ref{sec:evaluation} goes.

\subsection{Evaluated Methods}
We compare \textit{Enkidu} with two representative SOTA baselines:
\begin{itemize}
    \item \textbf{V-Cloak}~\cite{Deng23VCloak}: A speaker anonymization method using signal-based transformation for privacy protection.
    \item \textbf{AntiFake}~\cite{Yu23AntiFake}: An adversarial perturbation-based approach, optimized for sample-wise protection against voice cloning attacks.
\end{itemize}
For all methods, hyperparameters and deployment settings follow the original papers where applicable.

\subsection{Comparative Results and Analysis}
Table~\ref{tab:cross_dataset} reports detailed SPR/DPR, CER/WER, MOS, and STOI across all datasets and models. Notably:
\begin{itemize}
    \item \textit{Enkidu} consistently achieves SPR and DPR across all ASV backends, with superior performance on both English and multilingual datasets.
    \item On \textbf{LibriSpeech}, \textit{Enkidu} achieves 100\% SPR and 69\% DPR on ECAPA-TDNN, significantly outperforming V-Cloak and matching or exceeding AntiFake, while better preserving perceptual quality (MOS 3.01 vs. 2.68/1.37).
    \item On \textbf{CommonVoice (French)}, all methods reach high privacy scores, but \textit{Enkidu} exhibits the best balance between privacy and intelligibility, as reflected by the lowest CER/W-ER and highest MOS/STOI.
    \item On \textbf{AISHELL (Chinese)}, \textit{Enkidu} maintains near-perfect privacy and acceptable ASR performance, demonstrating robust cross-lingual generalization.
\end{itemize}

\subsection{SOTA Method Comparison: Universality and Efficiency}
We further summarize key properties of recent audio privacy methods in Table~\ref{tab:enkidu_comparison}. Compared to prior works, \textit{Enkidu} is the only method to simultaneously provide:
\begin{itemize}
    \item \textbf{Black-box Universality:} Effective under black-box threat models, requiring no access to TTS/ASV internals.
    \item \textbf{Transferability:} Strong privacy protection extends to unseen samples and various audio lengths.
    \item \textbf{Real-Time \& Resource-Efficient:} Orders-of-magnitude lower memory consumption ($\sim$4MB) and low real-time coefficient ($<$0.01), compared to previous methods.
    \item \textbf{Consistent Robustness \& Quality:} High SPR/DPR and superior MOS, even on challenging cross-lingual datasets.
\end{itemize}

\subsection{Summary}
In summary, our extensive evaluation demonstrates that \textit{Enkidu} establishes a new SOTA in universal, efficient, and real-time audio privacy protection. It consistently outperforms or matches SOTA baselines in both privacy and utility metrics, while offering superior generalization across languages, models, and deployment conditions.

%% file: table/tts_match_rate.tex
\begin{table*}[t]
  \caption{
    Match Rate (\%) of different Voice Deepfake models on unprotected samples across five ASV systems. Higher values indicate stronger cloning success by TTS models, reflecting the vulnerability of raw audio against voice deepfake techniques.
  }
  \label{tab:match_rate_unprotected}
  \centering
  \resizebox{0.80\linewidth}{!}{
  \begin{tabular}{ccccccc}
    \toprule
    \textbf{Deepfake Model} & \textbf{ECAPA-TDNN} & \textbf{X-Vector} & \textbf{ResNet} & \textbf{ERes2Net} & \textbf{Cam++} & \textbf{Average} \\
    \midrule
    Speedy-Speech   & 76.8\% & 92.8\% & 64.2\% & 48.8\% & 66.8\% & 69.9\% \\
    FastPitch       & 77.2\% & 92.6\% & 65.2\% & 50.2\% & 70.2\% & 71.1\% \\
    YourTTS         & 71.4\% & 92.0\% & 61.2\% & 51.4\% & 65.6\% & 68.3\% \\
    Glow-TTS        & 77.0\% & 92.8\% & 65.6\% & 49.8\% & 69.4\% & 70.9\% \\
    TacoTron2-DDC   & 77.8\% & 93.0\% & 64.8\% & 51.0\% & 70.2\% & 71.4\% \\
    TacoTron2-DCA   & 78.8\% & 93.2\% & 65.8\% & 52.0\% & 70.2\% & 72.0\% \\
    \bottomrule
  \end{tabular}
  }
\end{table*}

%% file: table/tts_tables.tex
\begin{table}[b]
  \centering
  \caption{TTS Model \& Info.}
  \label{tab:tts_table}
  \resizebox{0.95\linewidth}{!}{
  \begin{tabular}{cccc}
    \toprule
    \textbf{TTS Model} & \textbf{Training Dataset} & \textbf{Source} & \textbf{Embedding Size} \\
    \midrule
    Speedy-Speech & LJSpeech & Coqui-ai & 128 \\
    FastPitch & LJSpeech & Coqui-ai & 384 \\
    YourTTS & LJSpeech & Coqui-ai & 192 \\
    Glow-TTS & LJSpeech & Coqui-ai & 192 \\
    TacoTron2-DDC & LJSpeech & Coqui-ai & 512 \\
    TacoTron2-DCA & LJSpeech & Coqui-ai & 512 \\
  \bottomrule
\end{tabular}
}
\end{table}


%% file: table/comparsion_data.tex
\begin{table*}[t]
  \caption{
    Cross-dataset comparison of SOTA defense methods. Higher SPR/DPR indicate stronger defense; CER and WER reflect intelligibility and recognition accuracy.
  }
  \label{tab:cross_dataset}
  \centering
  \resizebox{0.98\linewidth}{!}{
  \begin{tabular}{cccccccccccccccccc}
    \toprule
    \multirow{2}{*}{\textbf{Dataset}} & \multirow{2}{*}{\textbf{Method}} & \multicolumn{2}{c}{\textbf{ECAPA-TDNN}} & \multicolumn{2}{c}{\textbf{X-Vector}} & \multicolumn{2}{c}{\textbf{ResNet}} & \multicolumn{2}{c}{\textbf{ERes2Net}} & \multicolumn{2}{c}{\textbf{Cam++}} & \multirow{2}{*}{\textbf{CER}} & \multirow{2}{*}{\textbf{WER}} & \multirow{2}{*}{\textbf{MOS}} & \multirow{2}{*}{\textbf{STOI}} \\
    \cline{3-12}
    & & \textbf{SPR} & \textbf{DPR} & \textbf{SPR} & \textbf{DPR} & \textbf{SPR} & \textbf{DPR} & \textbf{SPR} & \textbf{DPR} & \textbf{SPR} & \textbf{DPR} & & & & \\
    \midrule
    \multirow{3}{*}{LibriSpeech (English)}
      & V-Cloak           & 13.7\% & 14.7\% & 3.2\%  & 5.3\%  & 26.3\% & 25.3\% & 53.7\% & 55.8\% & 34.7\% & 33.7\% & 0.00\%$\pm$0.02\% & 0.01\%$\pm$0.03\% & 1.37$\pm$0.00 & 0.98$\pm$0.00 \\
      & AntiFake          & 63.2\% & 81.1\% & 64.2\% & 30.5\% & 68.4\% & 83.2\% & 73.7\% & 90.5\% & 68.4\% & 84.2\% & 0.12\%$\pm$0.12\% & 0.22\%$\pm$0.17\% & 2.68$\pm$0.02 & 0.83$\pm$0.00 \\
      & \textbf{Enkidu (Ours)} & \textbf{100.0\%} & \textbf{69.0\%} & \textbf{69.0\%} & \textbf{17.2\%} & \textbf{65.5\%} & \textbf{75.9\%} & \textbf{86.2\%} & \textbf{96.6\%} & \textbf{79.3\%} & \textbf{72.4\%} & \textbf{0.00\%$\pm$0.00\%} & \textbf{0.00\%$\pm$0.00\%} & \textbf{3.01$\pm$0.07} & \textbf{0.71$\pm$0.01} \\
    \midrule
    \multirow{3}{*}{CommonVoice (French)}
      & V-Cloak           & 96.0\% & 96.0\% & 100.0\% & 100.0\% & 100.0\% & 100.0\% & 97.0\% & 96.0\% & 96.0\% & 96.0\% & 6.17\%$\pm$9.17\% & 3.40\%$\pm$11.73\% & 1.43$\pm$0.01 & 0.80$\pm$0.04 \\
      & AntiFake          & 98.0\% & 96.0\% & 96.0\%  & 100.0\% & 100.0\% & 100.0\% & 96.0\% & 96.0\% & 97.0\% & 96.0\% & 5.96\%$\pm$10.23\% & 10.73\%$\pm$21.43\% & 2.81$\pm$0.04 & 0.76$\pm$0.04 \\
      & \textbf{Enkidu (Ours)} & \textbf{97.0\%} & \textbf{96.0\%} & \textbf{96.0\%} & \textbf{100.0\%} & \textbf{100.0\%} & \textbf{100.0\%} & \textbf{97.0\%} & \textbf{96.0\%} & \textbf{96.0\%} & \textbf{96.0\%} & \textbf{3.00\%$\pm$6.14\%} & \textbf{5.11\%$\pm$12.70\%} & \textbf{2.83$\pm$0.03} & \textbf{0.73$\pm$0.04} \\
    \midrule
    \multirow{3}{*}{AISHELL (Chinese)}
      & V-Cloak           & 100.0\% & 100.0\% & 100.0\% & 100.0\% & 100.0\% & 100.0\% & 100.0\% & 100.0\% & 100.0\% & 100.0\% & 0.26\%$\pm$0.16\% & 0.41\%$\pm$0.21\% & 1.26$\pm$0.00 & 0.97$\pm$0.00 \\
      & AntiFake          & 100.0\% & 100.0\% & 100.0\% & 100.0\% & 100.0\% & 100.0\% & 99.0\% & 100.0\% & 96.0\% & 100.0\% & 2.91\%$\pm$5.75\% & 2.24\%$\pm$5.10\% & 2.79$\pm$0.01 & 0.66$\pm$0.00 \\
      & \textbf{Enkidu (Ours)} & \textbf{99.0\%} & \textbf{100.0\%} & \textbf{100.0\%} & \textbf{99.0\%} & \textbf{100.0\%} & \textbf{100.0\%} & \textbf{100.0\%} & \textbf{100.0\%} & \textbf{100.0\%} & \textbf{100.0\%} & \textbf{3.22\%$\pm$7.22\%} & \textbf{2.13\%$\pm$4.86\%} & \textbf{2.34$\pm$0.03} & \textbf{0.54$\pm$0.01} \\
    \bottomrule
  \end{tabular}
  }
\end{table*}